\documentclass[fleqn,usenatbib]{mnras}

\usepackage{newtxtext,newtxmath}
\usepackage{comment}
\usepackage[export]{adjustbox}

\usepackage[T1]{fontenc}

\DeclareRobustCommand{\VAN}[3]{#2}
\let\VANthebibliography\thebibliography
\def\thebibliography{\DeclareRobustCommand{\VAN}[3]{##3}\VANthebibliography}

\usepackage{graphicx}	
\usepackage{amsmath}	
\usepackage[flushleft]{threeparttable}
\usepackage{booktabs,caption}

\def\M87{M87$^{\ast}$}
\def\m87{M87$^{\ast}$}
\def\sgra{Sgr A$^{\ast}$}

\def\MIseventeen{EHTC\,\M872017\,{I}}
\def\MIIseventeen{EHTC\,\M872017\,{II}}
\def\MIIIseventeen{EHTC\,\M872017\,{III}}
\def\MIVseventeen{EHTC\,\M872017\,{IV}}
\def\MVseventeen{EHTC\,\M872017\,{V}}
\def\MVIseventeen{EHTC\,\M872017\,{VI}}
\def\MVIIseventeen{EHTC\,\M872017\,{VII}}
\def\MVIIIseventeen{EHTC\,\M872017\,{VIII}}
\def\MIXseventeen{EHTC\,\M872017\,{IX}}
\def\MMWseventeen{EHTC\,MWL\,\M872017}
\def\MIeighteen{EHTC\,\M872017\,{I}}

\def\SIseventeen{EHTC\,\sgra2017\,{I}}
\def\SIIseventeen{EHTC\,\sgra2017\,{II}}
\def\SIIIseventeen{EHTC\,\sgra2017\,{III}}
\def\SIVseventeen{EHTC\,\sgra2017\,{IV}}
\def\SVseventeen{EHTC\,\sgra2017\,{V}}
\def\SVIseventeen{EHTC\,\sgra2017\,{VI}}
\def\SVIIseventeen{EHTC\,\sgra2017\,{VII}}
\def\SVIIIseventeen{EHTC\,\sgra2017\,{VIII}}

\defcitealias{M87EHTC2017I}{{\MIseventeen}}
\defcitealias{M87EHTC2017II}{{\MIIseventeen}}
\defcitealias{M87EHTC2017III}{{\MIIIseventeen}}
\defcitealias{M87EHTC2017IV}{{\MIVseventeen}}
\defcitealias{M87EHTC2017V}{{\MVseventeen}}
\defcitealias{M87EHTC2017VI}{{\MVIseventeen}}
\defcitealias{M87EHTC2017VII}{{\MVIIseventeen}}
\defcitealias{M87EHTC2017VIII}{{\MVIIIseventeen}}
\defcitealias{M87EHTC2017IX}{{\MIXseventeen}}
\defcitealias{M87EHTMWL2017}{{\MMWseventeen}}
\defcitealias{M87EHTC2018I}{{\MIeighteen}}
\defcitealias{SgrAEHTC2017I}{{\SIseventeen}}
\defcitealias{SgrAEHTC2017II}{{\SIIseventeen}}
\defcitealias{SgrAEHTC2017III}{{\SIIIseventeen}}
\defcitealias{SgrAEHTC2017IV}{{\SIVseventeen}}
\defcitealias{SgrAEHTC2017V}{{\SVseventeen}}
\defcitealias{SgrAEHTC2017VI}{{\SVIseventeen}}
\defcitealias{SgrAEHTC2017VII}{{\SVIIseventeen}}
\defcitealias{SgrAEHTC2017VIII}{{\SVIIIseventeen}}

\title[Variability in 2T GRMHD simulations]{Two-temperature treatments in magnetically arrested disk GRMHD simulations more accurately predict light curves of Sagittarius A$^\ast$}

\author[L.D.S. Salas et al.]{
L.D.S. Salas,$^{1}$\thanks{E-mail: l.d.sosapantasalas@uva.nl}
M.T.P. Liska,$^{2,3}$
S.B. Markoff,$^{1,4}$
K. Chatterjee,$^{5,6}$
G. Musoke,$^{7,1}$
O. Porth,$^{1}$
B. Ripperda,$^{7-10}$
\\
\textup{\Large{D. Yoon,$^{11}$,
W. Mulaudzi$^{1}$}}
\\
$^{1}$Anton Pannekoek Institute for Astronomy, University of Amsterdam, Science Park 904, 1098 XH Amsterdam, The Netherlands\\
$^{2}$Center for Relativistic Astrophysics, Georgia Institute of Technology, Howey Physics Bldg, 837 State St NW, Atlanta, GA, 30332, USA\\
$^{3}$Institute for Theory and Computation, Harvard University, 60 Garden Street, Cambridge, MA 02138, USA\\
$^{4}$Gravitation and Astroparticle Physics Amsterdam Institute, University of Amsterdam, Science Park 904, 1098 XH 195 196 Amsterdam, The Netherlands\\
$^{5}$Institute for Research in Electronics and Applied Physics, University of Maryland, 8279 Paint Branch Drive, College Park, MD 20742, USA\\
$^{6}$Black Hole Initiative at Harvard University, 20 Garden Street, Cambridge, MA 02138, USA\\
$^{7}$Canadian Institute for Theoretical Astrophysics, University of Toronto, 60 St. George Street, Toronto, ON M5S 3H8, Canada\\
$^{8}$Dunlap Institute for Astronomy and Astrophysics, University of Toronto, 50 St. George Street, Toronto, ON M5S 3H4, Canada\\
$^{9}$Department of Physics, University of Toronto, 60 St. George Street, Toronto, ON M5S 1A7, Canada\\
$^{10}$Perimeter Institute for Theoretical Physics, 31 Caroline Street North, Waterloo, ON N2L 2Y5, Canada\\
$^{11}$Information and Technology Services-Research Services, University of Iowa, Iowa 52242, Iowa City, IA 52242, USA\\
}

\date{Accepted XXX. Received YYY; in original form ZZZ}

\pubyear{2024}

\begin{document}
\label{firstpage}
\pagerange{\pageref{firstpage}--\pageref{lastpage}}
\maketitle


\begin{abstract}
The Event Horizon Telescope Collaboration (EHTC) observed the Galactic centre source Sagittarius A$^\ast$ (\sgra) and used emission models primarily based on single ion temperature (1T) general relativistic magnetohydrodynamic (GRMHD) simulations. This predicted emission is strongly dependent on a modelled prescription of the ion-to-electron temperature ratio. The most promising models are magnetically arrested disk (MAD) states. However, nearly all MAD models exhibit larger temporal fluctuations in radiative 230 GHz emission compared to observations. This limitation possibly stems from the fact that the actual temperature ratio depends on microphysical dissipation, radiative processes and other effects not captured in ideal fluid simulations. Therefore, we investigate the effects of two-temperature (2T) thermodynamics in MAD GRMHD simulations of \sgra, where the temperatures of both species are evolved. We find that the 230 GHz synchrotron flux variability more closely matches historical observations when we include the 2T treatment compared to 1T simulations. For the low accretion rates of \sgra, a common assumption is to neglect radiative cooling. However, we find that the radiative cooling of electrons—via synchrotron, inverse Compton, and bremsstrahlung processes—reduces the electron temperature in the inner disk, where the EHT observes, by about 10\%, which, in turn, decreases both the (sub)millimetre synchrotron flux and its temporal fluctuations compared to uncooled simulations. 

\end{abstract}






\section{Introduction}

\sgra\, was discovered as a bright compact radio source in the centre of the Milky Way galaxy \citep{Balick1974ApJ...194..265B, Ekers1975A&A....43..159E, Lo1975ApJ...202L..63L}. Subsequent observations provided compelling evidence that this object is a supermassive black hole (SMBH) by analysing its proper motion and the dynamics of individual stars in orbit around it \citep{Schodel2002Natur.419..694S,Ghez2003ApJ...586L.127G,Ghez2008ApJ...689.1044G,Gillessen2009ApJ...692.1075G,GRAVITYCollaboration2018A&A...615L..15G,Do2019Sci...365..664D,GRAVITY2019A&A...625L..10G}. More recently, the Event Horizon Telescope Collaboration (EHTC) presented direct evidence for the presence of this SMBH via imaging of near-event horizon regions \citepalias{SgrAEHTC2017I,SgrAEHTC2017VII}. The accretion flow of \sgra\, is collisionless based on the density inferred by, e.g., \citealt{Yuan2003ApJ...598..301Y,Dexter2010ApJ...717.1092D,Bower2019ApJ...881L...2B};  \citetalias{SgrAEHTC2017V,SgrAEHTC2017VIII}. In collisionless plasmas, the electron-ion collision timescale is much longer than the accretion timescale, electrons and ions are decoupled and not in thermal equilibrium, such that they may have two different temperatures (2T) \citep{Shapiro1976ApJ...204..187S,Rees1982Natur.295...17R,MahadevanQuataert1997ApJ...490..605M,Quataert1998ApJ...500..978Q}. These accretion flows are typically modelled with ideal general relativistic magnetohydrodynamics (GRMHD), which does not capture its collisionless and 2T nature. The most common GRMHD models treat the fluid as composed of single temperature ions (1T), where the electron density and temperature are not considered in the evolution equations \citep[e.g.][]{Gammie2003ApJ...589..444G,Tchekhovskoy2011MNRAS.418L..79T,Narayan2012MNRAS.426.3241N,McKinney2012MNRAS.423.3083M}. The (sub)millimetre emission is dominated by synchrotron radiation from relativistic electrons, making it crucial to accurately model electron thermodynamics. \cite{Ressler2015MNRAS.454.1848R,Ressler2017MNRAS.467.3604R} introduced a method to evolve the GRMHD equations describing a gas consisting of ions and electrons that share the same dynamics but have independent thermodynamical evolution. In this so-called 2T treatment, there is only an additional electron entropy equation, while the particle number and energy-momentum equations continue to assume a single fluid (ions). Therefore, we investigate how incorporating this 2T treatment and the commonly ignored electron radiative cooling impacts the predicted (sub)millimetre variability from simulations of \sgra.


Decades of observations of \sgra\, give very strong constraints on the (sub)millimetre variability, which is usually quantified with the modulation index $M_{3} \equiv \sigma_{3} / \mu_{3}$, where the standard deviation $\sigma_{3}$ and the mean $\mu_{3}$ are measured over three hour time intervals of the light curve. \cite{Wielgus2022ApJ...930L..19W} used the Atacama Large Millimeter/submillimeter Array (ALMA) and the Submillimeter Array (SMA) as individual interferometers during the EHT observations on 2017 April 5–11. They studied the light curves of \sgra\, at four frequencies bands between 213 and 229 GHz, with a minimum cadence of approximately 10 s. They reported that \sgra\, exhibited a low flux density of $2.4 \pm 0.2 \mathrm{Jy}$ and overall low variability, with non overlapping values $M_{3}=[0.024 - 0.051]$ across April 5–10. On April 11, the ALMA observations immediately followed an X-ray flare, with a corresponding enhanced variability $M_{3}=[0.084 - 0.117]$. The modulation index is consistent with other observations in 2005–2019 at frequencies around 230 GHz documented in various published works with longer cadence and lower number of collected data points \citep{Marrone2006JPhCS..54..354M,Marrone2008ApJ...682..373M,Yusef-Zadeh2009ApJ...706..348Y,Dexter2014MNRAS.442.2797D,Fazio2018ApJ...864...58F,Bower2018ApJ...868..101B,Witzel2021ApJ...917...73W,Iwata2020ApJ...892L..30I,Murchikova2021ApJ...920L...7M}.

1T GRMHD simulations include spherical accretion models \citep[e.g.][]{Ressler2021MNRAS.504.6076R,Lalakos2024ApJ...964...79L,Galishnikova2025ApJ...978..148G}, wid-fed models \citep{Ressler2020ApJ...896L...6R,Ressler2023MNRAS.521.4277R}, and torus-initialized models in the weakly magnetized (SANE) and magnetically arrested \citep[MAD; ][]{Bisnovatyi-Kogan1974Ap&SS..28...45B,Narayan2003PASJ...55L..69N} regimes. In the MAD regime the accretion is choked by the strong horizon penetrating magnetic field \citep[e.g.][]{Tchekhovskoy2011MNRAS.418L..79T,Porth2021MNRAS.502.2023P}. In these 1T simulations the ion-to-electron temperature\footnote{$T_i$ is the ion temperature and $T_e$ is the electron temperature.} ratio $T_i/T_e$ is determined using the so-called $R(\beta)$ prescriptions \citep[e.g.][]{Moscibrodzka2016A&A...586A..38M,Anantua2020MNRAS.493.1404A}, which is the main uncertainty in EHT modelling. None of the EHTC models of \sgra\, fully satisfy all the constraints drawn from multiwavelength observations at 86 GHz, 230 GHz, 2.2 $\mathrm{\mu m}$, and in the X-ray (for a detailed explanation, see \citetalias{SgrAEHTC2017II,SgrAEHTC2017V}). 230 GHz light curve variability at poses a particularly stringent challenge, as nearly all MAD models—and a significant portion of SANE models—exhibit greater variability ($M_3 \lesssim 0.5$) than seen in historical observations (\citealt{Wielgus2022ApJ...930L..19W} and references therein). On the other hand, the more realistic stellar wind-fed accretion models better predict the submillimetre variability in \sgra\ due to the comparatively lower levels of small-scale turbulence compared to SANE and MAD models \citep{Murchikova2022ApJ...932L..21M}. 
This variability problem could potentially be attributed to not modelling the evolution of $T_e$ when using the $R(\beta)$ prescriptions. In reality, $T_e$ is fundamentally influenced by microphysical plasma and radiation interactions, and does not depend trivially on $T_i$. A first-principles kinetic approach is required to model these collisionless effects \citep{Parfrey2019PhRvL.122c5101P,Crinquand2022PhRvL.129t5101C,Galishnikova2023PhRvL.130k5201G}. Nonetheless, it is possible to effectively model the electron thermodynamics of thermal electrons with 2T treatments in extended GRMHD simulations \citep[e.g.][]{Ressler2015MNRAS.454.1848R} and by including radiative effects \citep[e.g.][]{Sadowski2017MNRAS.466..705S,Chael2018MNRAS.478.5209C}.

The mechanisms of heating in collisionless plasma remain largely unconstrained. Typically used heating models include prescriptions for weakly collisional turbulent cascades (\citealt{Howes2010MNRAS.409L.104H}, H10; \citealt{Kawazura2019PNAS..116..771K}, K19) and magnetic reconnection mechanisms (\citealt{Rowan2017ApJ...850...29R}, R17; \citealt{Rowan2019ApJ...873....2R}, R19). These heating models have limited applicability, as reconnection and turbulence likely occur in diverse conditions across the disk and jet, making it improbable for a single, universally applied model to capture their complexity. \citet[][]{Sadowski2017MNRAS.466..705S,Ryan2015ApJ...807...31R, Ryan2017ApJ...844L..24R,Chael2018MNRAS.478.5209C,Chael2019MNRAS.486.2873C,Liska2024ApJ...966...47L} implemented the 2T treatment in a GR radiation MHD (GRRMHD) scheme with coupling between gas, radiation, magnetic ﬁelds and gravity. This 2T treatment does not capture realistic dissipative heating (see section~\ref{sec:2T}). When assuming that ions and electrons are heated through magnetic reconnection (R17 model), it was demonstrated that electrons are always cooler than ions and that electron heating is more uniform \citep{Chael2018MNRAS.478.5209C}. Additionally, the R17 model was favoured in a polarimetric parameter survey of \sgra, comparing H10, K19 and R17 \citep{Dexter2020MNRAS.494.4168D}. Because of the 2T nature of the accretion flow, a gas mixture consisting of relativistic electrons with an adiabatic index $\gamma_{e} = 4/3$ and non-relativistic ions with $\gamma_{i} = 5/3$ was found for the mass accretion rate of \sgra\, and R17 model \citep{Chael2018MNRAS.478.5209C,Liska2024ApJ...966...47L}. As a result, assuming a single fixed value for the gas adiabatic index could lead to inconsistencies between the total and individual temperatures and pressures. The adiabatic index characterizes the fluid response to compression, relating gas pressure $p_g$ and density $\rho$ via $p_g \propto \rho^\gamma$. Distinct adiabatic indices are important because electron thermodynamics govern observable emissions, while ion dynamics influence the bulk flow. Therefore, it is crucial to conduct a detailed comparison between the evolved $T_e$ from 2T GRMHD simulations with variable adiabatic indices and the $T_e$ calculated using the $R(\beta)$ prescription.

The estimated Eddington ratio\footnote{$\dot{M}$ is the mass accretion rate, $\dot{M}_{\mathrm{edd}}=4\pi GM/(c\kappa_{es}\eta_{\mathrm{NT}})$ is the Eddington accretion rate, $M$ is the mass of the black hole, $\kappa_{es}$ is the opacity due to electron scattering,
and $\eta_{\mathrm{NT}} = 0.178$ is the radiative efficiency of a thin accretion disk with black hole spin $a=0.9375$ \citep{NovikovThorne1973blho.conf..343N}. For \sgra\,, $\dot{M}_{\mathrm{edd}}\approx0.05 M_{\odot}/\mathrm{yr}.$} of \sgra$\,$ is  $f_{\mathrm{edd}}\equiv \dot{M}/\dot{M}_{\mathrm{edd}}\sim[1-100]\times10^{-8}$ 
(e.g. \citealt{Agol2000ApJ...538L.121A,Bower2003ApJ...588..331B,Marrone2007ApJ...654L..57M}). A common assumption for \sgra$\,$ is that radiative cooling and transport do not have a significant impact on the accretion dynamics and (sub)millimetre emission. In the advection-dominated accretion flow (ADAF) model, the radiative efficiency $\eta_{\mathrm{rad}}$ is less than 0.001 at all radii, meaning that less than $0.1\%$ of the available accretion energy is radiated \citep{Narayan1994ApJ...428L..13N,Narayan1995Natur.374..623N}. \cite{Fragile2009ApJ...693..771F,Dibi2012MNRAS.426.1928D} were pioneers in incorporating electron radiative cooling processes \citep{Esin:96}, including bremsstrahlung, synchrotron, and inverse Compton, into simulations. When $f_{\mathrm{edd}} > 10^{-7}$ for a 1T GRMHD SANE state, electron cooling lowers the scale-height of the accretion disk and the overall flux of the spectra from the submillimetre to the far-UV   \citep{Drappeau2013MNRAS.431.2872D,Yoon2020MNRAS.499.3178Y}. Similarly, variations in $T_i/T_e$ were identified when $f_{\mathrm{edd}} \gtrsim 10^{-7}$ in a 2T treatment \citep{Dihingia2023MNRAS.518..405D}. For a SANE state, radiation has a negligible effect on either the dynamics or the thermodynamics of the accreting gas when $f_{\mathrm{edd}}\sim 2\times 10^{-8}$ \citep{Sadowski2017MNRAS.466..705S}. Radiative cooling was found to lower $T_e$ in the inner regions of the accretion flow for $f_{\mathrm{edd}} > 10^{-6}$ in SANE states \citep{Ryan2015ApJ...807...31R, Ryan2017ApJ...844L..24R}. On the other hand, a MAD state has significantly higher level of magnetic flux and therefore higher $\eta_{\mathrm{rad}}$ due to more efficient synchrotron emission. When $f_{\mathrm{edd}} \approx 10^{-7}$, $\eta_{\mathrm{rad}}\approx0.03$ for a MAD state and $\eta_{\mathrm{rad}}\approx0.002$ for a SANE state \citep{Liska2024ApJ...966...47L}. Therefore, radiative cooling may be important for simulating \sgra\, if the accretion flow is in a MAD state.

In this paper, we study the variability of synchrotron radiative emission at [43-1360] GHz predicted from 2T GRRMHD simulations in a MAD state. We include variable adiabatic index \citep{Sadowski2017MNRAS.466..705S}, and radiative cooling of electrons (synchrotron, inverse Compton, and bremsstrahlung; \citealt{Esin1997ApJ...489..865E}) appropriate for \sgra. In our implementation, the cooling source terms subtract energy and momentum (see section~\ref{sec:2T}; \citealt{Fragile2009ApJ...693..771F,Dibi2012MNRAS.426.1928D}). Synchrotron emission dominates the radiative cooling rates and flux at [43-1360] GHz \citep{Yoon2020MNRAS.499.3178Y}. We also include Coulomb coupling that characterizes the transfer of energy between ions and electrons that occurs through collisions in a kinetic manner (see appendix~\ref{sec:Coulomb}; \citealt{Stepney1983MNRAS.204.1269S}).
Additionally, we assume that electron/ion heating occurs via magnetic reconnection (R17 model), and we do not account for anisotropic thermal conduction along magnetic field lines, as considered in \cite{Ressler2015MNRAS.454.1848R}. Recent studies have incorporated similar or more advanced physics at lower grid resolutions \citep{Chael2018MNRAS.478.5209C,Dexter2020MNRAS.494.4168D,Ressler2023MNRAS.521.4277R}. However, an analysis of variability that systematically adds one layer of physics at a time has been lacking. We quantify the variability of the total synchrotron flux density using the three-hour modulation index $M_3$. Our results show that the inclusion of 2T treatment, variable adiabatic index, and cooling systematically shifts the $M_3$ distribution towards lower variability values at 86, 228, 345, and 1360 GHz.

The structure of this paper is as follows: Section~\ref{sec:H-AMR} provides the 2T GRMHD equations and numerical setup for our simulations. The results of electron radiative cooling, heating, variable adiabatic index, and ray tracing are discussed in Section~\ref{sec:Results}. The conclusions of the study are presented in Section~\ref{sec:Conclusions}. 

\section{Simulations}
\label{sec:H-AMR}

\subsection{2T GRMHD equations}
\label{sec:2T}

We use geometrized units with gravitational constant, black hole mass, and speed of light $G=M=c=1$, and a factor of $1/\sqrt{4\pi}$ is absorbed in the definition of the magnetic field. The gravitational radius is $r_g \equiv GM/c^2$. Greek indices run through $[0,1,2,3]$ and Roman indices through $[1,2,3]$. The metric determinant is $g$. A semicolon ; is used to represent a covariant derivative. In the M1 closure scheme implemented in \cite{Sadowski2017MNRAS.466..705S,Chael2018MNRAS.478.5209C,Liska2024ApJ...966...47L}, radiation (approximated as a fluid) and gas are coupled through the exchange of four-momentum, a process known as radiative transport. We do not include the radiation energy-momentum equation, i.e., we neglect radiative transport. Instead, we solve the equations of GRMHD that comprise the particle number conservation equation:
\begin{equation}
     \left ( nu^{\mu} \right )_{;\mu} = 0\text{ ,}
     \label{eq:number_conservation_equation}
\end{equation}
the energy-momentum conservation equations:
\begin{equation}
     {T^{\mu}}_{\nu;\mu} = u_{\nu}q^{-}\text{ ,}
     \label{eq:energy_momentum_conservation_equation}
\end{equation}
and the Maxwell's equations:
\begin{equation}
     {F^{\ast\mu\nu}}_{;\nu} = 0\text{ .}
     \label{eq:Maxwell_equation}
\end{equation}

In ideal MHD, the dual of the electromagnetic ﬁeld
tensor is $F^{\ast\mu\nu}=b^{\mu}u^{\nu}-b^{\nu}u^{\mu}$ and the stress-energy tensor is:
\begin{equation}
    {T^{\mu}}_{\nu}\equiv \left (\rho+u_g+p_g+b^2 \right ) u^{\mu} u_{\nu}-b^{\mu} b_{\nu}+\left (\frac{1}{2}b^2+p_g \right ){\delta^{\mu}}_{\nu} \text{ ,}
	\label{eq:Tud}
\end{equation}
here $n$ is the particle number density, $\rho=n m_p$ is the rest mass density\footnote{Normalised to a maximum density $\rho _{\textup{max}}=1$ for the 1T simulation.}, $m_p$ is the proton mass, $u_g$ is the gas energy density, $p_g$ is the gas pressure, and $b^2/2$ is the magnetic energy density. The 4-velocity and the magnetic field 4-vector are $u^{\mu}$ and $b^{\mu}$ respectively. ${\delta^{\mu}}_{\nu}$ is the Kronecker delta.
The gas pressure is proportional to the gas temperature ($p_g\propto T_g $).  The magnetic pressure is $p_b=b^2/2$, where $b$ is the magnetic field strength in the frame co-moving with the fluid. $q^-$ is the radiative cooling rate of electrons (see section~\ref{sec:cooling}). We assume that proton cooling is insignificant, as proton synchrotron emission is negligible in the (sub)millimetre regime \citep[e.g.][]{SgrAEHTC2017V}. We consider a purely hydrogen fluid, so the relative mass abundances of hydrogen and helium are $X=1$ and $Y=0$. 
Therefore, the number densities are $n_{e}=n_{i}=n$.

Single fluid GRMHD assumes that both electrons and ions move at the same bulk velocity $u_e^{\mu}=u_i^{\mu}=u^{\mu}$, so their momenta are considered separately. The first law of thermodynamics determines the evolution of the species entropies,
\begin{equation}
    T_{e}\left ( n_es_eu^{\mu} \right )_{;\mu}=\delta _eq^{\textrm{v}}+q^C+q^-\text{ ,}
	\label{eq:se}
\end{equation}
\begin{equation}
    T_{i}\left ( n_is_iu^{\mu} \right )_{;\mu}=\delta _iq^{\textrm{v}}-q^C\text{ ,}
	\label{eq:si}
\end{equation}
where $s_{e,i}$ and $T_{e,i}$ are the entropy per particle and temperature of electrons and ions, respectively. $q^{\textrm{v}}=q^\mathrm{v}_e+q^\mathrm{v}_i$ is total dissipative heating rate (or "viscous" heating rate), $\delta _e=q^\mathrm{v}_e/q^{\textrm{v}}$ and $\delta _i=1-\delta_e$ the electron- and ion-to-overall heating ratio, and $q^C$ is the Coulomb coupling rate (energy exchange between electrons and ions).

The entropy per particle of an ideal gas with fixed adiabatic index $\gamma$ is $s \propto \ln (p/\rho^{\gamma})$. 
We follow the approach in \cite{Sadowski2017MNRAS.466..705S}, and use an approximate relativistic entropy formula,
\begin{equation}
    s_{e,i}=k_B\ln \left [ \frac{\Theta_{e,i}^{3/2}\left ( \Theta_{e,i}+2/5 \right )^{3/2}}{\rho_{e,i}} \right ]\text{ ,}
	\label{eq:8}
\end{equation}
which is analytically invertible for the dimensionless temperature $\Theta_{e,i}=k_BT_{e,i}/m_{e,i}c^2$, where $k_B$ is the Boltzmann constant. The ideal gas equation of state provides the means to determine the effective temperature of the gas $T_g$, which is a mix of electrons and ions,
\begin{equation}
    p_g=\left ( \gamma_g-1 \right )u_g=\frac{k_B\rho}{m_p}T_g\text{ .}
	\label{eq:9}
\end{equation}

The electron temperature $T_e$ is obtained from the entropy density equation \ref{eq:se}. Both species satisfy the equation of state,
\begin{equation}
    p_{e,i}=\left ( \gamma_{e,i}-1 \right )u_{e,i}=\frac{k_B\rho}{m_p}T_{e,i}\text{ ,}
	\label{eq:10}
\end{equation}

\begin{table*}
    \begin{threeparttable}
    	\centering
    	\caption{Setup parameters and simulation outcomes at quasi-steady state.}
    	\label{tab:Sims}
    	\begin{tabular}{ccccccc} 
    		\hline
    		Name & Cooling & Adiabatic index$^\textrm{a}$ & $\rho_s$(cgs)/$10^{-19}$ & $ F(\nu)^\textrm{b}/\mathrm{Jy}$ & $ f_{\mathrm{edd}}^\textrm{c}/10^{-7}$ & $t_{\mathrm{f}} (10^3 r_g/c)$ \\
    		\hline
    		1T & off & $\gamma_i = 5/3$ & - & - & - & 29.1\\
                2T & off & $\textrm{var}(\gamma)$ & 2.7 & 3.2 & 1.0 & 32.7\\
                2TC & on & $\textrm{var}(\gamma)$ & 2.7 & 2.7 & 1.0 & 29.1\\
    		2TC-8.3$\mathrm{Jy}$ & on & $\textrm{var}(\gamma)$ & 4 & 8.3 & 2.2 & 15.4\\
    		2TC-17.9$\mathrm{Jy}$ & on & $\textrm{var}(\gamma)$ & 10 & 17.9 & 4.2 & 21.2\\
    		2TC-93.8$\mathrm{Jy}$ & on & $\textrm{var}(\gamma)$ & $10^2$ & 93.8 & 38.6 & 21.5\\
    		2TC-62.0$\mathrm{Jy}$ & on & $\textrm{var}(\gamma)$ & $10^3$ & 62.0 & 365.1 & 21.1\\
    		\hline
            
    	\end{tabular}
        \begin{tablenotes}
          \small
          \item $^\textrm{a}$ $\textrm{var}(\gamma)$ represents variable adiabatic index \citep{Sadowski2017MNRAS.466..705S}. $^\textrm{b}$ Time-averaged synchrotron flux density at 228 GHz. $^\textrm{c}$ Time-averaged Eddington ratio measured at the event horizon. 
        \end{tablenotes}
    \end{threeparttable}
\end{table*}

The gas pressure/energy is the sum of the electron and ion pressures/energies, $p_g=p_e+p_i$ and $u_g=u_e+u_i$. Therefore, the temperatures must satisfy, $T_g= \left ( T_e +T_i\right )/2$, and the effective adiabatic index of the mixture is,
\begin{equation}
    \gamma_g=1+\frac{\left ( \gamma_{e}-1 \right )\left ( \gamma_{i}-1 \right )\left ( T_i/T_e+1 \right )}{\left ( T_i/T_e \right )\left ( \gamma_{e}-1 \right )+\left ( \gamma_{i}-1 \right )}\text{ ,}
	\label{eq:gamma_g}
\end{equation}
where $\gamma_{e,i}$ are the electron and ion adiabatic indices \citep{Narayan2011MNRAS.416.2193N,Sadowski2017MNRAS.466..705S},
\begin{equation}
    \gamma_{e,i}\approx \frac{10+20\Theta _{e,i}}{6+15\Theta _{e,i}}\text{ .}
	\label{eq:gamma_ei}
\end{equation}

We follow the method of \cite{Ressler2015MNRAS.454.1848R} to numerically identify the total dissipative heating $q^{\textrm{v}}$ by evolving the thermal entropies adiabatically over a time step $\Delta\tau$. By comparing the sum of the adiabatically evolved energy densities, $u_{i, \textrm{ad}}$ and $u_{e, \textrm{ad}}$, to the separately evolved total gas energy $u_g$, we estimate the dissipative heating in the total fluid as $q^{\textrm{v}}=(u_g - u_{i, \textrm{ad}} - u_{e, \textrm{ad}})/\Delta\tau$. The numerical implementation of mixing finite-sized fluid parcels into a single homogenised fluid within a cell artificially increases the entropy of the gas, which should not be treated as dissipation. As a result, this method loses accuracy in regions of the accretion flows with large discontinuities, such as those found in MAD states (for more details, see \citealt{Sadowski2017MNRAS.466..705S, Chael2025arXiv250112448C}). This method does not capture real dissipation. In reality, the physical processes that produce dissipation occur at scales much smaller than the simulation grid. We do not explicitly model the heating mechanisms via effective resistivity or viscosity. Instead, we assume that heating at the grid scale occurs through sub-grid magnetic reconnection \citep{Rowan2017ApJ...850...29R}. 

\subsubsection{Heating}



Previous studies on grid-scale dissipation have considered mainly two heating prescriptions: turbulent heating and magnetic reconnection. The non-relativistic turbulent heating model H10 was originally developed for solar wind observations and is based on calculations of a turbulent cascade in a weakly collisional plasma with $\sigma_i \ll 1$ \citep{Howes2010MNRAS.409L.104H}. The relativistic turbulent heating model K19 exhibits quantitatively similar behaviour to the H10 model \citep{Kawazura2019PNAS..116..771K}. For \sgra, assuming H10 results in a negligible difference in electron temperatures in MAD models compared to K19 \citep{Dexter2020MNRAS.494.4168D}.

We adopt the heating of electrons and ions resulting from magnetic reconnection with zero guide field \citep{Rowan2017ApJ...850...29R,Chael2018MNRAS.478.5209C}. This model is based on particle-in-cell simulations with physical mass ratio $m_i/m_e = 1836$ and total magnetisation for relativistic particles, that properly accounts for relativistic inertia $\sigma_{\textrm{w}}=b^2/\textrm{w}\geqslant 0.03$. The electron-to-overall heating ratio $\delta_e$ is
\begin{equation}\label{eq:deltae_Rowan}
	\delta_e = 0.5\exp\left [ \frac{\beta_i/\beta_{i,\mathrm{max}} -1}{0.8+\sigma_{\textrm{w}}^{0.5}} \right ]\text{ ,}
\end{equation}
where the enthalpy density per unit volume is $\textrm{w} = \rho_i c^2+\gamma_{e}u_e+\gamma_{i}u_i$. The ratio of ion thermal pressure to magnetic pressure is $\beta_i=2k\rho_i T_i/\left (m_ib^2\right )$, where $\beta_i\leqslant \beta_{i,\:\mathrm{max}}=1/\left ( 4\sigma_{\textrm{w}}\right )$. The heating prescription including the effect of guide ﬁelds by \citet{Rowan2019ApJ...873....2R} is quantitatively similar to zero guide field case Eq. \ref{eq:deltae_Rowan}. Unlike H10, Eq. \ref{eq:deltae_Rowan} varies less rapidly with $\beta_i$, is always non-zero and never allocates more than half of the heat to electrons.

\subsection{Numerical setup}

We perform the GRMHD simulations using the H-AMR code \citep{Liska2022ApJS..263...26L}, which builds on the HARM2D code \citep{Gammie2003ApJ...589..444G,Noble2006ApJ...641..626N}. H-AMR employs GPU acceleration within a hybrid CUDA-OpenMP-MPI framework, integrating adaptive mesh refinement (AMR) and local adaptive time-stepping (LAT) to facilitate efficient and scalable simulations. We use spherical Kerr–Schild coordinates, where $t$, $r$, $\theta$ and $\phi$ are the temporal, radial, polar and azimuthal coordinates, respectively. The number of cells in the radial, polar and azimuthal directions are $N_{r}\times N_{\theta} \times N_{\phi} = 520 \times 224 \times 224$. The radial domain is $r=\left [ 1.2 - 2000 \right ]r_g$. We use outflow boundary conditions in $r$, transmissive boundary conditions in $\theta$, and periodic boundary conditions in $\phi$, as described in \citet{Liska2022ApJS..263...26L}. In all runs the disk is initialised using a torus in hydrostatic equilibrium \citep{Fishbone1976RelativisticHoles} around a Kerr black hole with dimensionless spin $a=0.9375$. The initial inner edge of the torus is located at $r=20 r_g$ and the pressure maximum at $r=41 r_g$. The torus is threaded with a single poloidal magnetic field loop, defined by the $\phi$-component of the vector potential $A_{\phi} \propto  \textup{max}\left [ \rho/\rho _{\textup{max}}\left ( r/r_\textup{in} \right )^3 \sin ^3\theta \exp\left ( -r/400 \right )-0.2 , 0\right ]$, and normalised to obtain $\beta=p_{g}/p_{b}=100$ at pressure maxima, specifically to reach a MAD state. The following floor and ceiling values are employed; the rest-mass density floor is $\rho _{\textup{fl}}=\mathrm{max}[b^2/25,10^{-7}r^{-2},10^{-20}]$, the gas energy density floor is $u _{g,\textup{fl}}=\mathrm{max}[b^2/750,10^{-9}r^{-26/9},10^{-20}]$ and the magnetisation ceiling is $\sigma _{\textup{max}}=25$ where $\sigma=b^2/\rho$. The disk is initialised with the same energy density perturbation $u_g(1 + 0.04(\textrm{rand([0, 1])} - 0.5))$, where $\textrm{rand([0, 1])}$ is a random deviate between 0 and 1. 

For the 1T simulation, we assume a non-relativistic gas with a constant adiabatic index of 5/3. Table~\ref{tab:Sims} summarises the main simulation setup parameters, and results of 228 GHz flux density and Eddington ratio. The 1T simulation is scale-free, whereas the 2TC simulations require a density scale $\rho_s$ to compute the physical radiative cooling rates. To obtain the appropriate density scale, we performed an iterative process: running multiple 2TC simulations, performing ray-tracing, and verifying the time-averaged flux density at 228 GHz. The 2TC simulation (without extra labels) matches the 228 GHz flux density 2.7 Jy \citepalias{SgrAEHTC2017II}. The 2TC-x.x$\mathrm{Jy}$ simulations were initialized with higher density scales, resulting in higher accretion rates and increased 228 GHz synchrotron flux densities, $F(\nu)>8\mathrm{Jy}$. A higher plasma density increases opacity, thereby reducing synchrotron flux variability. Therefore, we compare the variability of synchrotron radiative emission only between simulations 1T, 2T and 2TC with $2.7\mathrm{Jy}$. 2TC follows the same initial setup as 2T but with cooling enabled. The values reported in Table~\ref{tab:Sims} and Section~\ref{sec:Results} for simulation 2T are obtained using the same density scale as in simulation 2TC. The primary simulations (1T, 2T, 2TC) are evolved for a time of approximately $3\times10^4r_g/c$, with a cadence of $10r_g/c$, equivalent to 3.5 minutes for \sgra. 


\section{Results}
\label{sec:Results}

In order to determine quantitative differences between the simulations we compare various fluid parameters. The averaged profile of a variable $X$ is calculated by integrating over $\theta$ and $\phi$. The accretion flow is taken to be the region that satisfies magnetisation $\sigma=b^2/\rho < 1$. Additionally, we include a density weight to give more relevance to regions of the disk with higher density and therefore higher thermal synchrotron emissivity $j_{\nu} = j_{\nu}(\rho,B,T_e)$,  \footnote{Note the difference in notation, $\left \langle  \right \rangle$ without any sub/super script is time average.}
\begin{equation}
    \left \langle X \right \rangle_{\rho[\theta,\phi]}^{\textrm{disk}} = \frac{\iint X \rho \left ( \sigma < 1 \right )\sqrt{-g}d\theta d\phi}{\iint \rho  \left ( \sigma < 1 \right ) \sqrt{-g}d\theta d\phi} \text{ .}
	\label{eq:AverageDisk}
\end{equation}

The fluxes are defined as follows. The mass accretion rate $\dot{M}$ is given by:
\begin{equation}
    \dot{M}\equiv -\iint \rho u^r\sqrt{-g}d\theta d\phi \text{ ,}
	\label{eq:Mdot}
\end{equation}
the energy flux $\dot{E}$ is given by:
\begin{equation}
    \dot{E}\equiv -\iint {T^{r}}_{t}\sqrt{-g}d\theta d\phi \text{ ,}
	\label{eq:Edot}
\end{equation}
the magnetic flux is defined as:
\begin{equation}
   \Phi\equiv \frac{\sqrt{4\pi}}{2}\iint \left | B^r \right |\sqrt{-g}d\theta d\phi
	\label{eq:Phi} \text{ .}
\end{equation}

For $r\lesssim 100r_g$, there is an inflow equilibrium in the disk so $\dot{M}_{\textrm{disk}}$ is approximately constant. The disk properties still depend on the initial conditions of the Fishbone-Moncrief torus for radii larger than $100 r_g$. To study MAD simulations it is convenient to use the normalised magnetic flux, $\varphi\equiv \Phi/\sqrt{ \left \langle \dot{M} \right \rangle}$, known as the “MAD parameter” which, for spin $a=0.9375$ and torus scale height $H/R \approx 0.3$, has the critical value $\varphi _{max}\approx 40-50$ (when using cgs units in the ratio; \citealt{tchekhovskoy2012general}). For SANE models, $\varphi _{max}\approx 7$ \citep{Chatterjee2022ApJ...941...30C}. $H$ and $R$ are the full height and cylindrical radius of the disk, respectively. The scale height is defined geometrically as $H/R\equiv \left \langle \left | \theta - \pi/2 \right | \right \rangle_{\rho[\theta,\phi]}^{\textrm{disk}}$. In general, subscripts indicate where in the radial domain the variables are analysed, e.g. $X_{BH}$ at the event horizon and $X_{10r_g}$ at $10r_g$.

\subsection{Radiative cooling}
\label{sec:cooling}

The total cooling rate for an optically thin gas is computed from the cooling function, 
\begin{equation}\label{eq:coolrate}
	q^- = \eta_{\rm br,C}\, q_{\rm br}^{-} + \eta_{\rm Sync,C}\, q_{\rm Sync}^{-}\,,
\end{equation}
where $q_{\rm br}^{-}$ and $q_{\rm Sync}^{-}$ are the bremsstrahlung and synchrotron cooling rates, respectively. We find that Bremsstrahlung cooling is negligible, consistent with \cite{Yoon2020MNRAS.499.3178Y}. $\eta_{\rm br,C}$ and $\eta_{\rm Sync,C}$ are the Compton enhancement factors for bremsstrahlung and synchrotron, respectively. These factors are the average energy gain of the photon in an assumption of single scattering \citep{Esin:96}. A detailed description of the cooling rates is available in \cite{Yoon2020MNRAS.499.3178Y}.

\begin{figure}
	\includegraphics[width=\columnwidth]{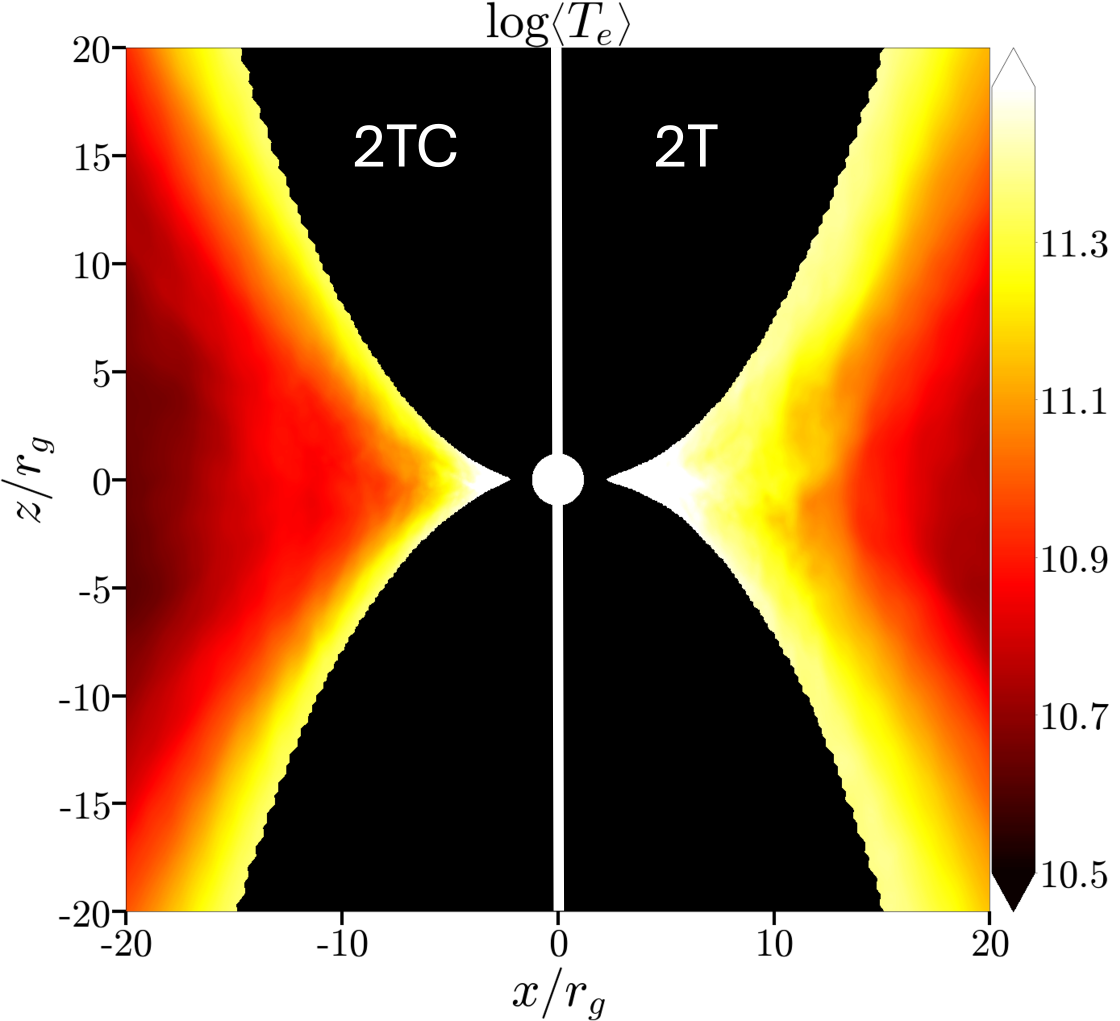}
        \caption{Cross sections in the $xz$-plane at $y = 0$ showing (log) electron temperature in Kelvin, averaged between $[16-29]\times 10^3 r_g/c$. 2T with cooling (left) and without cooling (right). Radiative cooling lowers the electron temperature in the inner accretion disk. We cover the jet spine region ($\sigma>1$) with a black screen.}
    \label{fig:Te}
\end{figure}

\begin{figure}
	\includegraphics[width=\columnwidth]{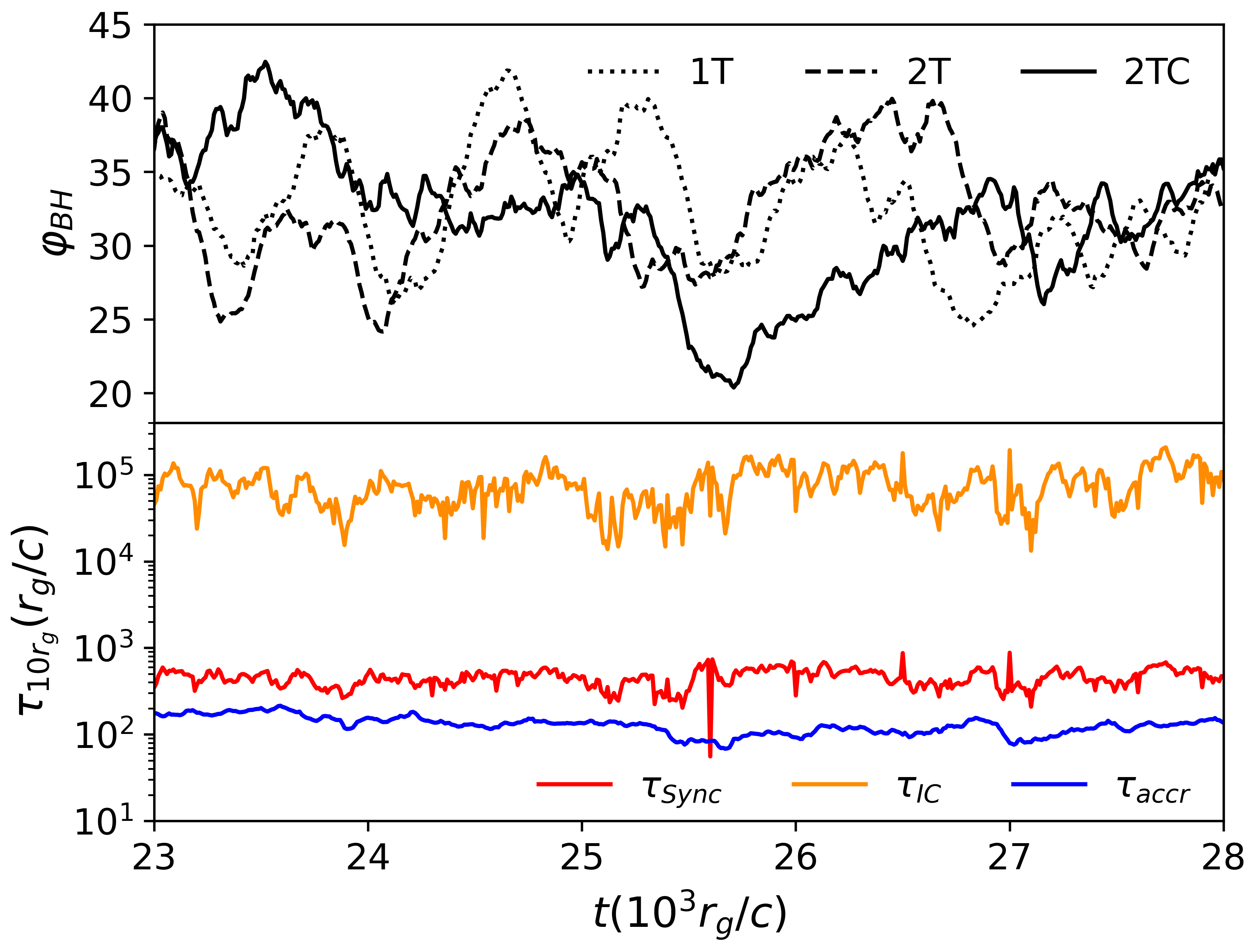}
    \caption{Normalised magnetic flux (top panel) and various timescales (bottom panel) as functions of time. The magnetic flux saturates after $t=5\times 10^3 r_g/c$. Line conventions are as follows: 1T—dotted, 2T—dashed, 2TC—solid. Timescales for synchrotron emission ($\tau_{\textup{Sync}}$), Comptonization ($\tau_{\textup{IC}}$), and accretion ($\tau_{\textup{accr}}$) at $10r_g$ highlight the relative significance of these processes.}
    \label{fig:disk_vs_t}
\end{figure}

\begin{figure}
	\includegraphics[width=\columnwidth]{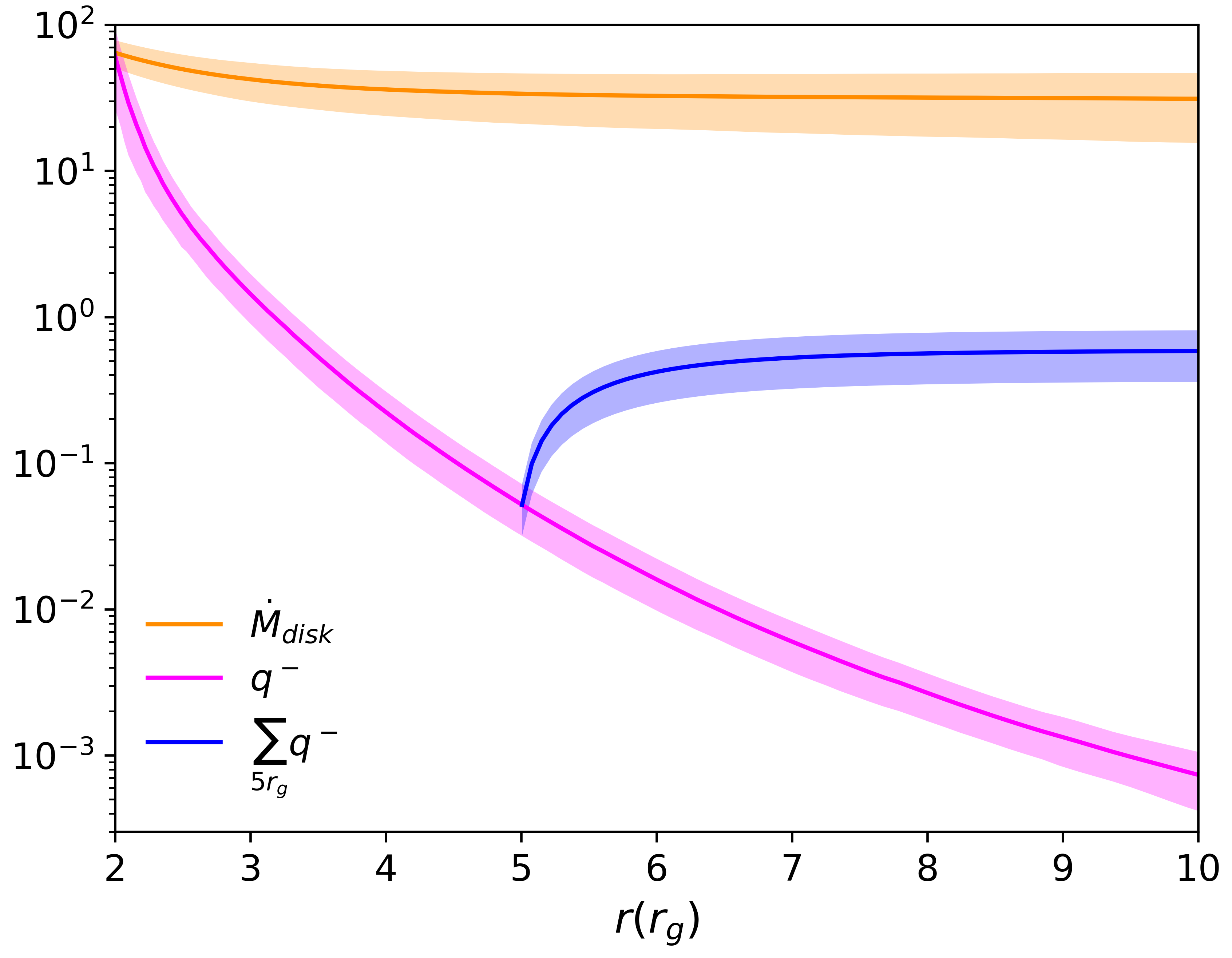}
    \caption{Disc averaged profiles of the total cooling rate $q^{-}$, and mass accretion rate $\dot{M}_{disk}$, averaged between $[23-28]\times 10^3 r_g/c$. Shaded regions depict the range of variation within one standard deviation. The estimated radiative efficiency is $\eta_{\mathrm{rad}} \approx 0.02 \pm 0.1$.}
    \label{fig:eta_rad_q}
\end{figure}

Fig.~\ref{fig:Te} shows a comparison of $T_e$ for simulations 2T and 2TC. Even for the very low accretion rate relevant for \sgra, $f_{\mathrm{edd}}\approx10^{-7}$, we find that radiative cooling still lowers $T_e$ in the inner disk and the average synchrotron flux by $10\%$ (see Table~\ref{tab:Sims}). The importance of the radiative losses is tested by computing the radiative cooling timescale $\tau = u_e/q^-$ and compare it to the dynamical accretion timescale. We calculate the synchrotron timescale as:
\begin{equation}
    \tau_{\textup{Sync}}\equiv \frac{\left \langle u_e \right \rangle_{\rho[\theta,\phi]}^{\textrm{disk}}}{\left \langle q_{\rm Sync}^{-} \right \rangle_{\rho[\theta,\phi]}^{\textrm{disk}}}
	\label{eq:tau_Sync} \text{ ,}
\end{equation}
the Comptonization timescale as:
\begin{equation}
    \tau_{\textup{IC}}\equiv \frac{\left \langle u_e \right \rangle_{\rho[\theta,\phi]}^{\textrm{disk}}}{\left \langle (\eta_{\rm Sync,C}-1) q_{\rm Sync}^{-} \right \rangle_{\rho[\theta,\phi]}^{\textrm{disk}}}
	\label{eq:tau_IC} \text{ ,}
\end{equation}
and the accretion timescale as:
\begin{equation}
    \tau_{\textup{accr}}\equiv \frac{\iint \rho r\sqrt{-g}d\theta d\phi}{\left \langle \dot{M}_{10r_g}\right \rangle}
	\label{eq:tau_accr} \text{ .}
\end{equation}


MAD states are characterised by large fluctuations caused by episodic magnetic flux eruptions. Once magnetic flux becomes over-saturated, magnetic reconnection occurs near the black hole, ejecting low-density, highly magnetised flux tubes into the disk \citep[e.g.][]{Porth2021MNRAS.502.2023P,Chatterjee2021MNRAS.507.5281C,Ripperda2022ApJ...924L..32R}. Fig.~\ref{fig:disk_vs_t} presents time series between $[23-28]\times 10^3 r_g/c$ of the normalised magnetic flux and the previously defined timescales. Across all simulations, the average flux $\left\langle \varphi \right\rangle \approx 35$ reveals distinct signatures of several magnetic flux eruption events. Our analysis indicates that the timescales, ranked by relevance, are as follows: synchrotron $  \left< \tau_{\textup{Sync}} \right>/ \left< \tau_{\textup{accr}}\right>\sim 3$, and Comptonization $  \left< \tau_{\textup{IC}}\right>/ \left< \tau_{\textup{accr}}\right>\sim 500$. Therefore, we find that synchrotron emission dominates over inverse Compton, consistent with \citet{Yoon2020MNRAS.499.3178Y,Liska2024ApJ...966...47L}.

In 2T GRRMHD simulations, \citet{Liska2024ApJ...966...47L} demonstrated that magnetic flux significantly influences radiative efficiency, primarily attributed to more efficient synchrotron emission in MAD models at low accretion rates (see also \citealt{Ryan2017ApJ...844L..24R,Dexter2021ApJ...919L..20D}). At $f_{\mathrm{edd}} \approx 10^{-7}$, $\eta_{\mathrm{rad}} \approx 0.002$ and $T_i/T_e \approx 10$ at $10r_g$ in the disk for SANE models, while for MAD models, $\eta_{\mathrm{rad}} \approx 0.03$ and $T_i/T_e \approx 3$ at $10r_g$ in the disk \citep{Liska2024ApJ...966...47L}. Their more self-consistent calculation of radiative efficiency incorporates the radiation stress-energy tensor in the M1 scheme. In our simulations, which include only radiative cooling, we estimate the radiative efficiency $\eta_{\mathrm{rad}} \approx \eta_{\mathrm{NT}}  \left( \tau_{\textup{accr}}/ \tau_{\textup{cool}}\right)_{10r_g} \approx 0.05\pm0.1$, where $\tau_{\textup{cool}}=(1/\tau_{\textup{Sync}}+1/\tau_{\textup{IC}})^{-1}$. Alternatively, using the cooling rate $\eta_{\mathrm{rad}}\approx  \sum_{5r_g}^{200r_g} q^{-}/\left \langle \dot{M}_{5r_g}\right \rangle\approx0.02\pm0.01$ (see Fig.~\ref{fig:eta_rad_q}). 


\subsection{Heating and variable adiabatic index}

\begin{figure*}
	\includegraphics[width=\textwidth]{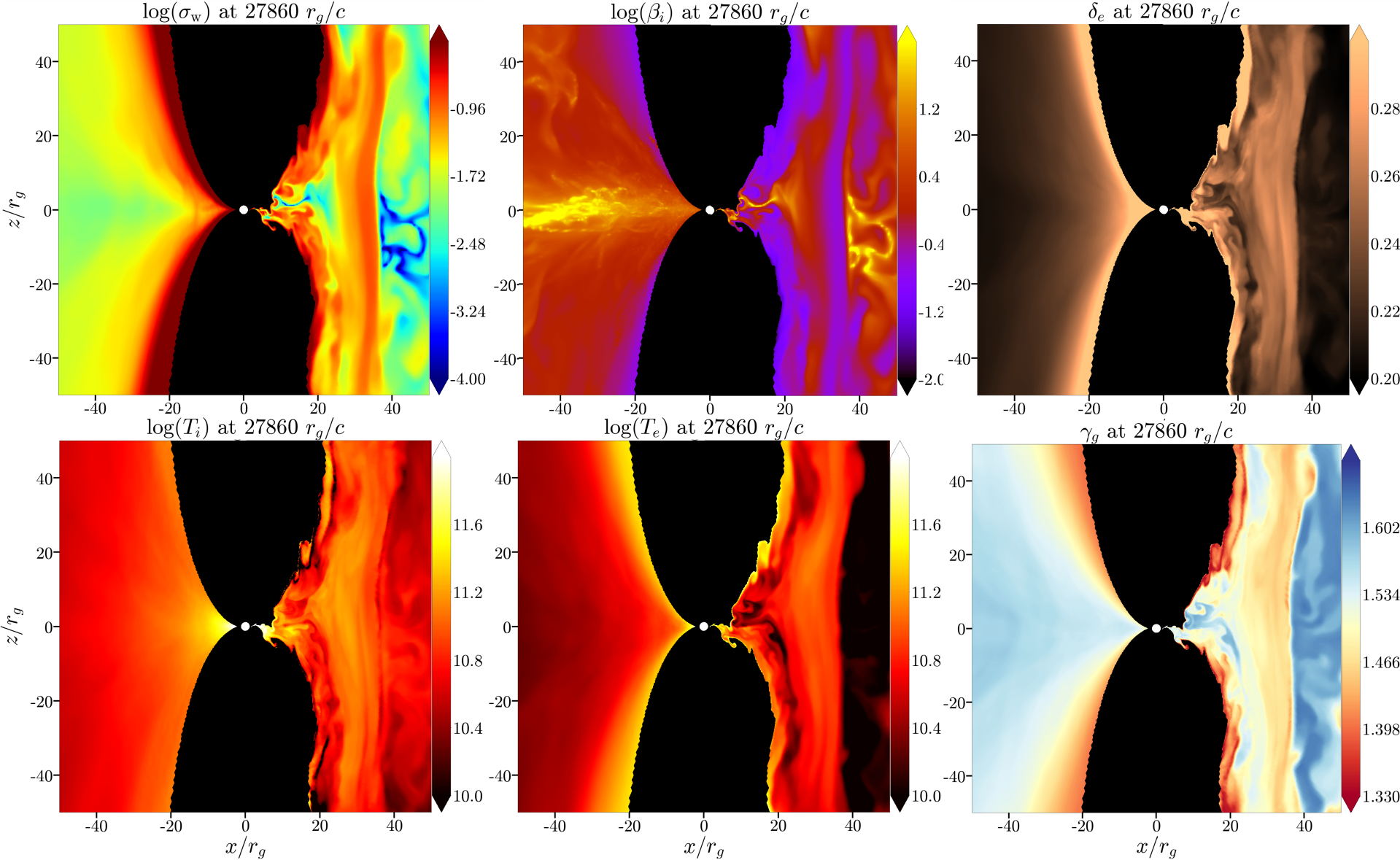}
    \caption{Cross sections in the $xz$-plane at $y = 0$ during a episode of magnetic flux eruption in simulation 2TC. First row: (log) total magnetisation $\sigma_{\textrm{w}}$ (left panel), (log) ion plasma beta $\beta_i$ (middle panel) and electron-to-overall heating ratio $\delta _e$ (right panel). Second row: (log) ion temperature in Kelvin $T_i$ (left), (log) electron temperature $T_e$ in Kelvin (middle) and adiabatic index of gas mixture (right). The left hemispheres show the time-averaged values between $[16-29]\times 10^3 r_g/c$ including flux eruptions, while the right hemispheres present a snapshot at $27860r_g/c$. We cover the jet spine region ($\sigma>1$) with a black screen. Electron heating is particularly prominent in the jet-disk interface and flux tubes, where magnetic pressure is high (high $\sigma_{\textrm{w}}$ and low $\beta_i$). The evolved $T_e$ surpasses $T_i$ in the jet-disk interface. The time-averaged $\gamma_g\approx1.55$ at $20r_g$. During a flux eruption, the temperatures in the jet-disk interface depend on the magnetisation ceiling as $T \propto \sigma_{\textup{max}}$  \citep{Ripperda2022ApJ...924L..32R}.}
    \label{fig:2TC_vertical}
\end{figure*}

Fig.~\ref{fig:2TC_vertical} shows the correlation between the magnetisation ($\sigma_{\textrm{w}}$), ion plasma beta ($\beta_i$), and the electron-to-overall heating ratio $\delta_e$ determined by Eq. \ref{eq:deltae_Rowan}. In general, $\delta_e$ determines the temperature and adiabatic index of both species and the gas mixture (see Eq. \ref{eq:se}, \ref{eq:si}, \ref{eq:gamma_g}, and \ref{eq:gamma_ei}). By assuming heating through magnetic reconnection, we observe the increase of $T_e$ more prominently in regions with high $\sigma_{\textrm{w}}$ and low $\beta_i$. In those regions, the adiabatic index of the electron/ion mixture, determined by $T_e$ and $T_i$, approaches 4/3, consistent with the plasma becoming relativistic as a result of heating. The 1T simulation employs a fixed value of 5/3. During an eruption in MAD states, the expulsion of magnetic flux goes through the magnetic reconnection of field lines in a current sheet in the equatorial plane. Highly magnetised plasma from the jet spine region ($\sigma=b^2/\rho>1$) supplies matter to the current sheet, with $T \propto \sigma_{\textup{max}}$, and the reconnection exhaust deposits this hot plasma in the jet-disk interface \citep{Ripperda2022ApJ...924L..32R}.

In the first-principles kinetic simulations of \citep{Rowan2017ApJ...850...29R}, magnetic energy is transformed into heat or kinetic energy when magnetic fields reconnect. This process accelerates and heats particles, with electron heating being particularly prominent in regions where magnetic pressure is high, like in the disk-jet boundary and flux tubes during episodes of magnetic flux eruption events. In GRRMHD SANE simulations of \sgra, \citet{Chael2018MNRAS.478.5209C} demonstrated that electrons are preferentially heated in the polar outflows with the H10 prescription \citep[see also e.g.][]{Sadowski2017MNRAS.466..705S,Ressler2017MNRAS.467.3604R}, whereas electrons are heated by nearly the same fraction everywhere in the accretion flow with the R17 prescription. Additionally, the models assuming the R17 heating model show variability in the 230 GHz light-curve more consistent with the level observed from \sgra\, (\citealt{Wielgus2022ApJ...930L..19W} and references therein).




%

\begin{figure}
	\includegraphics[width=\columnwidth]{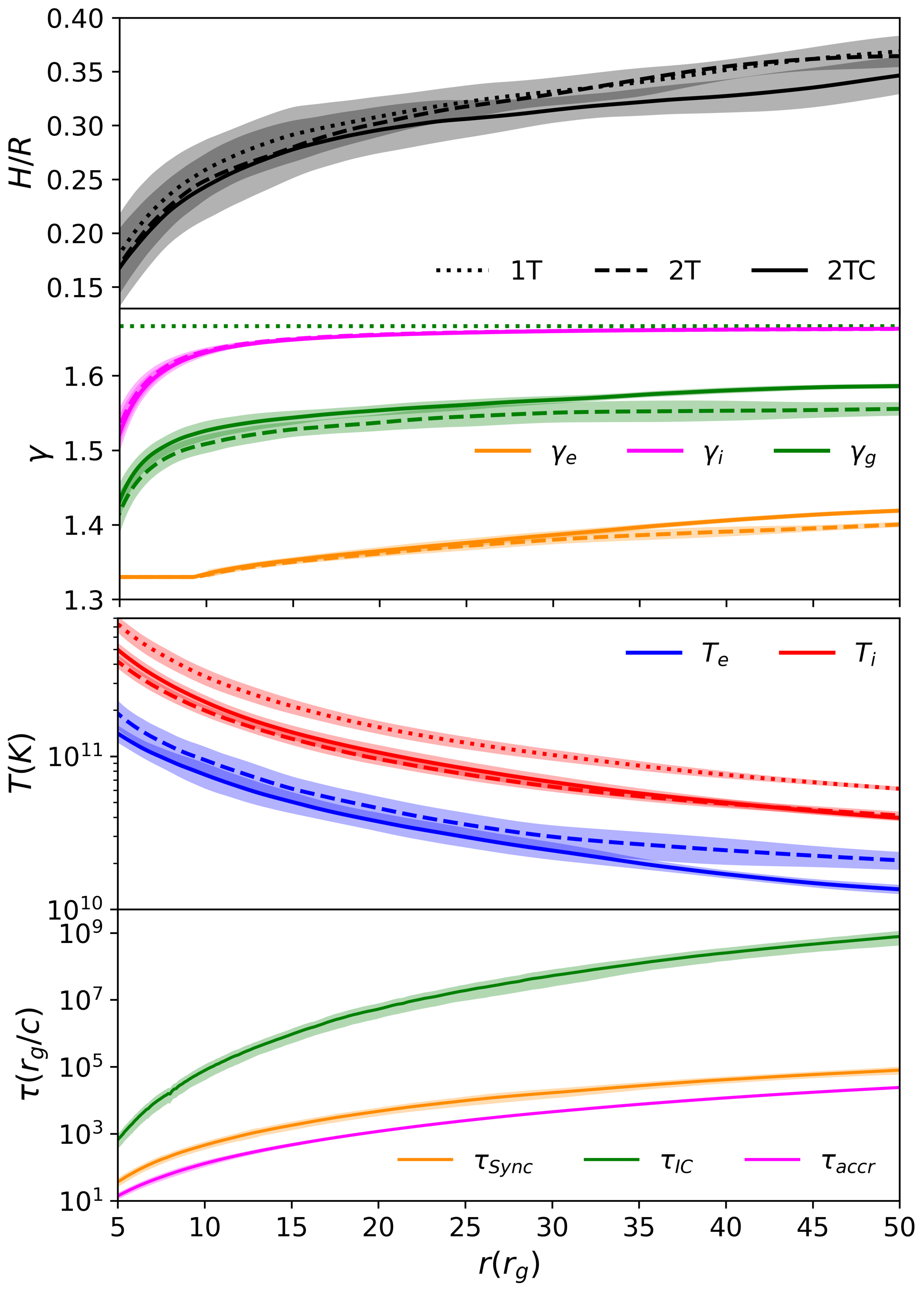}
    \caption{Disc averaged profiles of variables as a function of radius, averaged between $[22-29]\times 10^3 r_g/c$. First panel: density scale-height. Second panel: adiabatic indices. Third panel: electron and ion temperatures. Fourth panel: timescales at $27860 r_g/c$. Line conventions are as follows: 1T—dotted, 2T—dashed, 2TC—solid. Shaded regions depict the range of variation within one standard deviation. Inflow equilibrium is better converged in the final part of the simulation. Despite cooling lowers $T_e$, the thickness of the disk at $<15r_g$ remains unchanged because ion and magnetic pressures dominate over electron pressure. Lower adiabatic index values correspond to reduced pressure and temperature.}
    \label{fig:disk_vs_r}
\end{figure}

Fig.~\ref{fig:disk_vs_r} shows radial profiles of the scale-height, adiabatic indices, electron and ion temperatures, and the synchrotron, Comptonization, and accretion timescales (see section~\ref{sec:cooling}). Our findings reveal that the scale-height in 1T simulations is larger compared to the 2T and 2TC simulations, but within the one standard deviation range. The higher adiabatic index in the 1T simulation results in a higher $T_i$ and a larger scale-height. Comparing the 2T and 2TC simulations, we find that radiative cooling reduces $T_e$ by approximately $15\%$ within $50r_g$, with the most significant drop occurring between $[30-50]r_g$. This cooling effect also lowers the values of $\gamma_e$ and $\gamma_g$. We find that while the radial profile of the timescale ratio $\tau_{\textup{Sync}}/\tau_{\textup{accr}}$ remains relatively constant, $\tau_{\textup{IC}}/\tau_{\textup{accr}}$ increases significantly with increasing radius.

We observe a temperature ratio of $ \left< T_i/T_e \right>\approx 3$ and $ \left< T_e \right>\sim 10^{11} \text{K}$, both time averaged between $[16-29]\times 10^3 r_g/c$ and measured at $10r_g$, consistent with those reported in 2T simulations \citep[e.g.][]{Liska2024ApJ...966...47L}, and in spectral fitting modelling of 1T simulations where the ion-to-electron temperature ratio is a model parameter \citep[e.g.][]{Moscibrodzka2009ApJ...706..497M,Drappeau2013MNRAS.431.2872D}. \citet{Moscibrodzka2009ApJ...706..497M} found that models with $T_i/T_e = 1$ were inconsistent with the submillimeter spectral slope, while those with $T_i/T_e = 3$ and $10$ provided a better match to spectral and VLBI observations. Furthermore, with the inclusion of radiative cooling \citep{Dibi2012MNRAS.426.1928D}, \citet{Drappeau2013MNRAS.431.2872D} identified an optimal spectral fit for models with $T_i/T_e = 3$.

\subsection{Ray Tracing}

The ray-tracing of the GRMHD simulations is performed by solving the general-relativistic radiative transfer (GRRT) equations using the BHOSS code \citep{Younsi2012A&A...545A..13Y,Younsi2016PhRvD..94h4025Y}. Radiative processes are calculated using synchrotron emission and absorption, which are sufficient for imaging and light curve analysis across radio and submillimetre frequencies. In the submillimetre regime, synchrotron emission dominates the spectrum (\cite{Yoon2020MNRAS.499.3178Y};\citetalias{SgrAEHTC2017V}). The 2.2 $\mu \mathrm{m}$ emission is usually dominated by synchrotron, and the X-ray can be dominated by either Compton scattering or bremsstrahlung \citepalias{SgrAEHTC2017V}. To match the average synchrotron flux density of \sgra\, at 2.7 Jy at 228 GHz \citepalias{SgrAEHTC2017II}, the density scale is adjusted in the GRRT post-processing of the 1T simulation. For the 2T models, the density scale is directly obtained from the GRMHD simulations (see Table~\ref{tab:Sims}).

Based on the mean values of \sgra\, reported by \cite{Do2019Sci...365..664D} and \cite{GRAVITY2019A&A...625L..10G}, we assume the mass $M=4.14\times10^6M_{\odot}$ (same as in the 2T simulations) and distance $D=8.127\mathrm{kpc}$, where $M_{\odot}$ is the solar mass \citepalias{SgrAEHTC2017V}. Since there is no definitive evidence of a jet in \sgra\,, determining the exact orientation of the source relative to our line of sight remains challenging. Nevertheless, the model comparison of the EHTC suggests that high inclination angles ($i > 50^{\circ}$) are unlikely \citepalias{SgrAEHTC2017V}. We analyse the 1T simulation considering six inclination angles, $i=[10, 30, 50, 130, 150, 170]^{\circ}$, and the 2T simulations with the preferred value $i=30^{\circ}$ that passes most total intensity and polarimetric constraints \citepalias{SgrAEHTC2017V,SgrAEHTC2017VIII}. We find that $M_3$ distributions do not significantly depend on the inclination angle.

In the 1T simulation, the strength of electron–proton coupling is unknown, yet $T_e$ is crucial to calculate the radio synchrotron spectra assuming a thermal electron distribution function. Similarly as in \citet{Moscibrodzka2016A&A...586A..38M,moscibrodzka2017MNRAS.468.2214M}, we calculate $T_e$ using the parametrization with respect to the plasma beta ($\beta \equiv p_g/p_b$),
\begin{equation}
    \frac{T_i}{T_e}\equiv R_{\textup{high}}\frac{\beta ^2}{\beta ^2+1}+R_{\textup{low}}\frac{1}{\beta ^2+1} \text{ .}
	\label{eq:T_p/T_e}
\end{equation}

This $R(\beta)$ model was initially developed phenomenologically for SANE simulations and may not be suitable for MAD states, particularly in predicting submillimetre light curves. In regions of the accretion flow where $\beta \gtrsim 1$ or $\lesssim 10^{-2}$, the electron temperature is approximately independent of $\beta$. Consequently, $T_e$ fluctuates approximately as much as $T_i$ (see section~\ref{sec:Fluctuations_Te_Ti}).

For \sgra, electron distribution functions with approximately equal proton and electron temperatures ($R_{\textup{high}} = 1$) are unlikely based on the 2017 data observed by the EHT, and the two most promising models have $R_{\textup{high}} = 160$ \citepalias{SgrAEHTC2017V,SgrAEHTC2017VIII}, so that the electrons are much colder in the disk and hotter in the jet. However, 2T GRMHD simulations with H10, K19 and R17 heating prescriptions have consistently shown that time averaged $T_i/T_e\approx[3-10]$ in the accretion disk \citep[e.g.][]{Chael2018MNRAS.478.5209C,Mizuno2021MNRAS.506..741M,Dihingia2023MNRAS.518..405D,Liska2024ApJ...966...47L}. The sole value of $R_{\textup{low}}=1$ explored by the EHTC could potentially skew the high values of $R_{\textup{high}}$. When $f_{\mathrm{edd}} \approx 10^{-6}$ and $a=0.94$, \citet{Mizuno2021MNRAS.506..741M} found that MAD models using the R17 heating prescription show only a slight reduction in 230 GHz light-curve variability compared to predictions from 1T simulations using the $R(\beta)$ prescription. \cite{Moscibrodzka2024arXiv241206492M} demonstrated that 2T MAD models with K19 heating exhibit less variability at 228 GHz compared to $R(\beta)$ models, even when using a constant gas adiabatic index of $\gamma = 13/9$. Additionally, $M_3$ increases with black hole spin and slightly decreases when physics of non-thermal electrons are included. Moreover, their resolved images most closely resemble 1T models with $T_i/T_e = 10$ in both linear and circular polarisation.

\begin{figure*}
	\includegraphics[width=\textwidth]{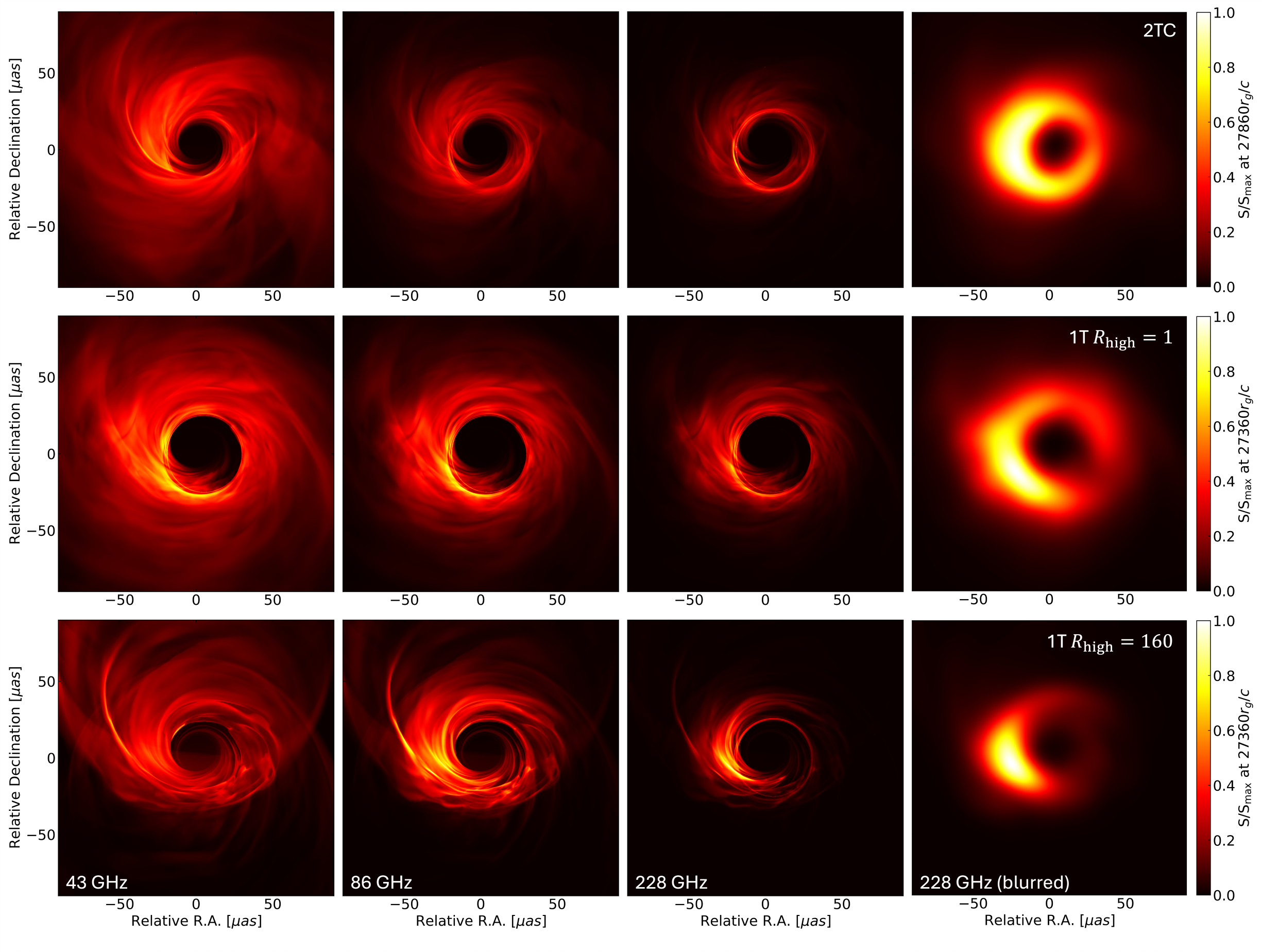}
    \caption{Ray-traced images at inclination $30^{\circ}$. Top row: 2TC simulation. Middle row: 1T simulation for  $R_{\mathrm{high}}=1$. Bottom row: 1T simulation for  $R_{\mathrm{high}}=160$. From left to right: 43, 86, 228 GHz (unblurred), and 228 GHz (blurred with a Gaussian kernel with $\textup{FWHM}=15\mu \mathrm{as}$ to simulate the resolution of the EHT). The mildly relativistic outflow in the disk-jet boundary is more visible for $R_{\mathrm{high}}=160$. The accretion flow transitions from optically thick at 43-86 GHz to optically thin at 228-345-1360 GHz. Therefore, the synchrotron radiation variability increases with frequency as shown in Table~\ref{tab:M3_freq}. Images from the 2T simulation are not included here, as they are nearly identical to those from the 2TC simulation.}
    \label{fig:BHshadow}
\end{figure*}

We explore five values $R_{\textup{high}}=[1,10,20,40,160]$. We find that the evolved $T_e$ surpasses $T_i$ in the jet-disk interface (see Fig.~\ref{fig:2TC_vertical}), whereas the $T_e$ calculated using the $R(\beta)$ prescription invariably remains lower than $T_i$. Fig.~\ref{fig:BHshadow} shows a visual comparison of the ray-traced images at inclination $i=30^{\circ}$ for simulations 2TC and 1T at different frequencies. Images from the 2T simulation are not included in Fig.~\ref{fig:BHshadow}, as they are nearly identical to those from the 2TC simulation. The density scales to match the time-averaged flux density of 2.7 Jy are $\rho_{\mathrm{scale}}=2.6$ and $15.4 \,(\mathrm{g\,cm^{-3}})$ for $R_{\mathrm{high}}=1$ and $160$, respectively. As $R_{\mathrm{high}}$ increases, emission from the disk ($\beta>1$) diminishes while emission from the rapidly moving outflows ($\beta<1$) intensifies. This effect results in increased synchrotron flux variability, as shown in Table~\ref{tab:M3_1T}.

Table~\ref{tab:M3_1T} also presents the jet power $P_{\mathrm{jet}}=\dot{M}_{\mathrm{jet}}-\dot{E}_{\mathrm{jet}}$, averaged over the region between $50$ and $500 r_g$, where the jet is identified based on the Bernoulli parameter $Be = -\bar{h} u_t>1.02$ where $\bar{h}=(\rho+u_g+p_g)/\rho$ is the specific gas enthalpy and $u_t$ is the time component of the inverse four-velocity \citep{Davelaar2018GeneralA}. The density scale derived from ray-tracing the 1T simulations is used to calculate the jet power in cgs units. For comparison, the jet power in the 2TC and 2T simulations is $(1.4 \pm 0.4) \times 10^{38} \mathrm{erg/s}$ and $(1.3 \pm 0.3) \times 10^{38} \mathrm{erg/s}$, respectively, which matches with the jet power in the 1T simulation for $R_{\textup{high}}$ between 1 and 10.

\begin{table}
    \centering
    \caption{Mean and standard deviation values of $M_3$ for the 1T simulation. $M_3$ increases with $R_{\textup{high}}$ but does not significantly depend on the inclination angle.}
    \label{tab:M3_1T}
    \begin{tabular}{ccccc} 
        \hline
        $R_{\textup{high}}$ & $i$ & $M_3$ & $\left \langle f_{\mathrm{edd}} \right \rangle/10^{-8}$ & $\left \langle P_{\mathrm{jet}} \right \rangle/10^{38}\mathrm{erg/s}$\\
        \hline
        160 & 10 & 0.26 $\pm$ 0.09 & 39.2 & 5.9 $\pm$ 1.5 \\
        160 & 30 & 0.26 $\pm$ 0.08 & 39.9 & 6.6 $\pm$ 1.7 \\
        160 & 50 & 0.27 $\pm$ 0.07 & 40.4 & 6.2 $\pm$ 1.6 \\
        160 & 130 & 0.27 $\pm$ 0.07 & 40.8 & 6.2 $\pm$ 1.6 \\
        160 & 150 & 0.27 $\pm$ 0.08 & 40.2 & 6.1 $\pm$ 1.6 \\
        160 & 170 & 0.26 $\pm$ 0.09 & 39.3 & 6.0 $\pm$ 1.0 \\
        40 & 30 & 0.24 $\pm$ 0.08 & 19.8 & 3.0 $\pm$ 0.8 \\
        20 & 30 & 0.23 $\pm$ 0.09 & 15.7 & 2.3 $\pm$ 0.6 \\
        10 & 30 & 0.21 $\pm$ 0.09 & 11.2 & 1.7 $\pm$ 0.4 \\
        1 & 10 & 0.18 $\pm$ 0.08 & 6.9 & 1.1 $\pm$ 0.3 \\
        1 & 30 & 0.17 $\pm$ 0.08 & 6.8 & 1.1 $\pm$ 0.3 \\
        1 & 50 & 0.17 $\pm$ 0.08 & 6.4 & 1.1 $\pm$ 0.3 \\
        1 & 130 & 0.18 $\pm$ 0.08 & 6.8 & 1.1 $\pm$ 0.3 \\
        1 & 150 & 0.18 $\pm$ 0.08 & 6.9 & 1.1 $\pm$ 0.3 \\
        1 & 170 & 0.18 $\pm$ 0.08 & 6.9 & 1.1 $\pm$ 0.3 \\
        \hline
    \end{tabular}
\end{table}

\begin{table}
    \centering
    \caption{Comparison of $M_3$ values at different frequencies for an inclination of $30^{\circ}$. 
    As frequency increases, $M_3$ also increases due to the accretion flow becoming more optically thin. The inclusion of 2T treatment, variable adiabatic index, and radiative cooling of electrons reduces $M_3$.}
    \label{tab:M3_freq}
    \begin{tabular}{clcc} 
        \hline
         $\nu(\textup{GHz})$ & Level of physics & $M_3$ \\
        \hline
        43 & \textup{1T} & 0.08 $\pm$ 0.04\\
        43 & \textup{1T}+\textup{var}($\gamma$) & 0.11 $\pm$ 0.04\\
        43 & \textup{1T}+\textup{var}($\gamma$)+\textup{Cool} & 0.09 $\pm$ 0.04\\
        43 & \textup{2T}+\textup{var}($\gamma$) & 0.06 $\pm$ 0.03\\
        43 & \textup{2T}+\textup{var}($\gamma$)+\textup{Cool} & 0.05 $\pm$ 0.02\\
        \hline
        86 & \textup{1T} & 0.12 $\pm$ 0.05\\
        86 & \textup{1T}+\textup{var}($\gamma$) & 0.15 $\pm$ 0.07\\
        86 & \textup{1T}+\textup{var}($\gamma$)+\textup{Cool} & 0.12 $\pm$ 0.05\\
        86 & \textup{2T}+\textup{var}($\gamma$) & 0.09 $\pm$ 0.05\\
        86 & \textup{2T}+\textup{var}($\gamma$)+\textup{Cool} & 0.07 $\pm$ 0.04\\
        \hline
        228 & \textup{1T} & 0.23 $\pm$ 0.09\\
        228 & \textup{1T}+\textup{var}($\gamma$) & 0.21 $\pm$ 0.09\\
        228 & \textup{1T}+\textup{var}($\gamma$)+\textup{Cool} & 0.16 $\pm$ 0.07\\
        228 & \textup{2T}+\textup{var}($\gamma$) & 0.14 $\pm$ 0.08\\
        228 & \textup{2T}+\textup{var}($\gamma$)+\textup{Cool} & 0.12 $\pm$ 0.05\\
        \hline
        345 & \textup{1T} & 0.26 $\pm$ 0.10\\
        345 & \textup{1T}+\textup{var}($\gamma$) & 0.22 $\pm$ 0.10\\
        345 & \textup{1T}+\textup{var}($\gamma$)+\textup{Cool} & 0.17 $\pm$ 0.07\\
        345 & \textup{2T}+\textup{var}($\gamma$) & 0.16 $\pm$ 0.09\\
        345 & \textup{2T}+\textup{var}($\gamma$)+\textup{Cool} & 0.13 $\pm$ 0.06\\
        \hline
        1360 & \textup{1T} & 0.32 $\pm$ 0.11\\
        1360 & \textup{1T}+\textup{var}($\gamma$) & 0.27 $\pm$ 0.11\\
        1360 & \textup{1T}+\textup{var}($\gamma$)+\textup{Cool} & 0.21 $\pm$ 0.08\\
        1360 & \textup{2T}+\textup{var}($\gamma$) & 0.21 $\pm$ 0.11\\
        1360 & \textup{2T}+\textup{var}($\gamma$)+\textup{Cool} & 0.16 $\pm$ 0.07\\
        \hline
    \end{tabular}
\end{table}

\begin{figure*}
	\includegraphics[width=\textwidth]{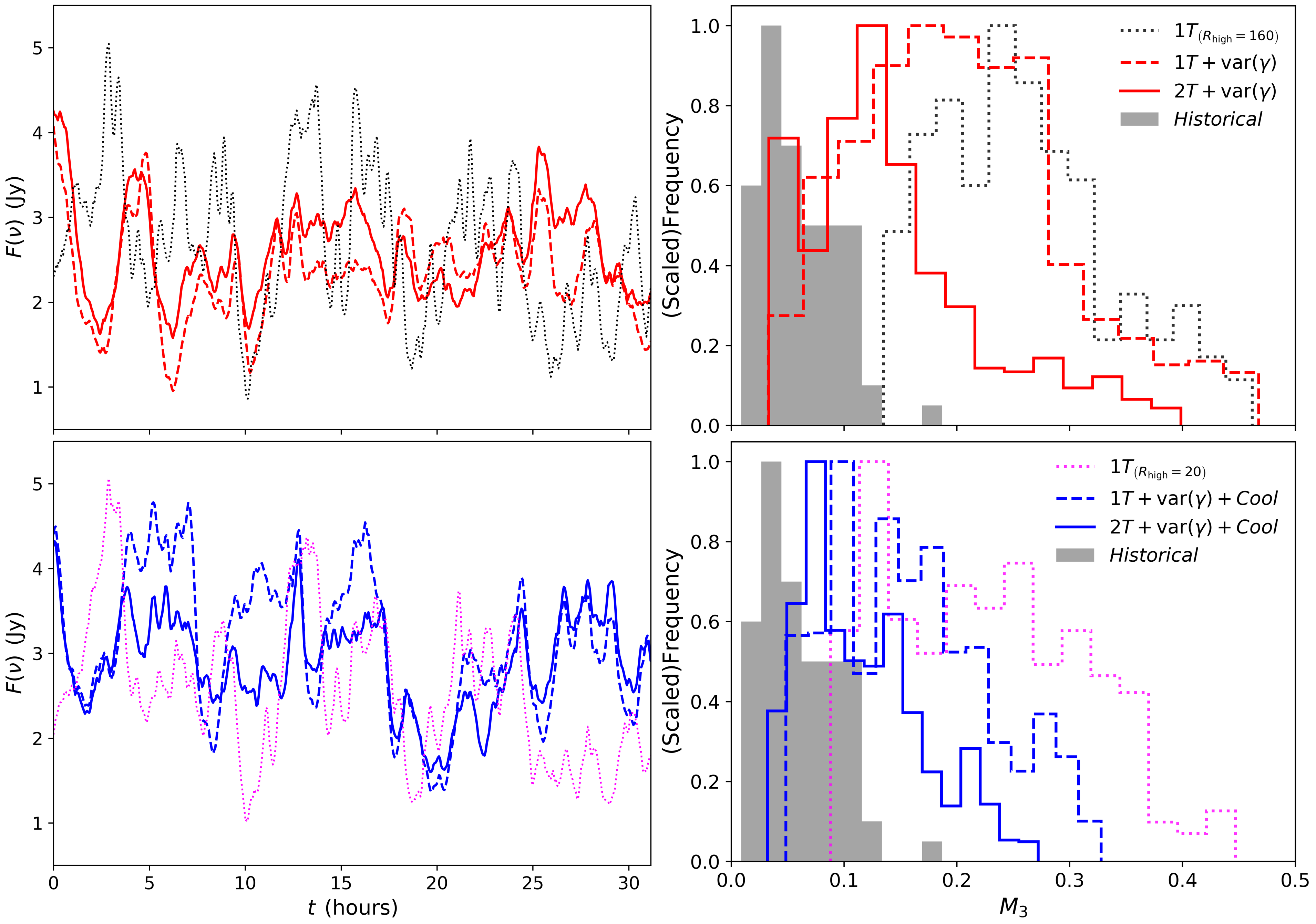}
    \caption{Light curves at 228 GHz (left panels) and $M_3$ distributions (right panels) at an inclination of $30^{\circ}$. The 1T simulations are shown with dotted lines for $R_{\mathrm{high}}=$ 20 (magenta) and 160 (black). The 2T simulations are represented without cooling (red) and with cooling (blue). Solid lines indicate when the evolved $T_e$ is directly taken from the 2T GRMHD simulations, while dashed lines indicate when the evolved $T_e$ is ignored and instead calculated in post-process using the $R(\beta)$ prescription. Historical $M_3$ observations are shown in grey (\citealt{Wielgus2022ApJ...930L..19W} and references therein). The 2T treatment with variable adiabatic index and the inclusion of radiative cooling brings the models closer to historical variability levels. Additionally, $M_3$ is lower when using the evolved $T_e$, indicating that the $T_e$ derived from the $R(\beta)$ prescription exhibits greater fluctuations.}
    \label{fig:Light-curve_M3}
\end{figure*}

We obtain distributions of the modulation index $M_3$ measured over three-hour intervals. Fig.~\ref{fig:Light-curve_M3} shows the 228 GHz light curves and $M_3$ distributions\footnote{The EHT observes in four frequency bands centred at 213.1 GHz (band 1), 215.1 GHz (band 2), 227.1 GHz (band 3), and 229.1 GHz (band 4). The $M_3$ values at 228 GHz and 230 GHz are indistinguishable in our simulations.} for $i=30^{\circ}$. Values of $M_3$ for the 1T simulation with different inclination angles and $R_{\textup{high}}$ are presented in Table~\ref{tab:M3_1T}. A comparison of $M_3$ at different frequencies for 1T, 2T and 2TC simulations is presented in Table~\ref{tab:M3_freq}. We find that the 2T treatment, variable adiabatic index and radiative cooling shifts the $M_3$ distribution towards lower variability values. However, historical observations of \sgra\,(\citealt{Wielgus2022ApJ...930L..19W} and references therein) still show lower $M_3$ values than 2T simulations.


From the 2T and 2TC simulations, we extract two light curve predictions per simulation: one using the evolved $T_e$ directly from the simulation, and another where the evolved $T_e$ is ignored and instead calculated in post-processing with the $R(\beta)$ prescription. We find that $M_3$ is lower when using the evolved $T_e$, suggesting that the $T_e$ derived from the $R(\beta)$ prescription exhibits greater fluctuations (see Fig~\ref{fig:Light-curve_M3}).

\subsection{Fluctuations in electron and ion temperatures}
\label{sec:Fluctuations_Te_Ti}

In the $R(\beta)$ prescriptions used in 1T simulations, $T_e$ fluctuates as much as $T_i$ in regions with $\beta \gtrsim 1$ or $\lesssim 10^{-2}$ where $T_e \approx T_i/R$ \citep{Moscibrodzka2016A&A...586A..38M}. However, that is not the case in a 2T treatment with different adiabatic indices for electrons and ions. Thermodynamically, a 2T treatment with $\gamma_e\neq\gamma_i$ is expected to have an impact in the fluctuation of $T_e$ \citetext{private communication, Gammie, 2024}. From the first law of thermodynamics, the convective derivatives of $T_e$ and $T_i$ are:
\begin{equation}\label{eq:DlnT_eDt}
	\frac{D\ln{T_e}}{Dt} = \frac{\gamma_e-1}{kn_eT_e}q_e+\frac{\gamma_e-1}{\tau_{\textrm{comp}}}\text{ ,}
\end{equation}
\begin{equation}\label{eq:DlnT_iDt}
	\frac{D\ln{T_i}}{Dt} = \frac{\gamma_i-1}{kn_iT_i}q_i+\frac{\gamma_i-1}{\tau_{\textrm{comp}}}\text{ ,}
\end{equation}
where the heating/cooling rate per unit volume $q=q^{\textrm{v}}+q^{C}+q^{-}$ includes ion-electron energy exchange $q^{C}$, dissipation heating rate $q^{\textrm{v}}$ and cooling rate $q^{-}$. For each species $q^{\textrm{v}}_e=\delta_eq^{\textrm{v}}$ and $q^{\textrm{v}}_i=\delta_iq^{\textrm{v}}$, with $\delta_e+\delta_i=1$ and the total dissipation rate $q^{\textrm{v}}=(u_e+u_i)/\tau_{\textrm{diss}}$. We consider only radiative cooling of electrons, i.e. $q^{-}_e=u_e/\tau_{\textrm{cool}}$ and $q^{-}_i=0$. $\tau_{\textrm{comp}}$, $\tau_{\textrm{diss}}$, $\tau_{\textrm{cool}}$ are the timescales of compression, dissipation, and cooling, respectively.

From our simulations $\gamma_e\approx4/3$, $\gamma_i\approx5/3$, $T_i/T_e\approx3$, $\delta_e\approx[0.2-0.4]$, and $\delta_i\approx[0.8-0.6]$. Analysing the adiabatic process case $q=0$, from Eq.~(\ref{eq:DlnT_eDt}) and (\ref{eq:DlnT_iDt}), $T_e$ fluctuates approximately half as much as $T_i$,
\begin{equation}\label{eq:DlnT_eDt_ad}
	\frac{D\ln{T_e}}{Dt} = \frac{\gamma_e-1}{\gamma_i-1}\frac{D\ln{T_i}}{Dt}\approx\frac{1}{2}\frac{D\ln{T_i}}{Dt}\text{ ,}
\end{equation}
in the incompressible case $\tau_{\textrm{comp}}\rightarrow \infty$ with only heating,  
\begin{equation}\label{eq:DlnT_eDt_heat}
	\frac{D\ln{T_e}}{Dt} = \frac{\gamma_e-1}{\gamma_i-1}\frac{T_i}{T_e}\frac{\delta_e}{\delta_i}\frac{D\ln{T_i}}{Dt}\approx[0.4-1]\frac{D\ln{T_i}}{Dt}\text{ ,}
\end{equation}
and the addition of cooling can reduce the fluctuations of $T_e$ as compared to $T_i$,
\begin{equation}\label{eq:DlnT_eDt_heat-cool}
	\frac{D\ln{T_e}}{Dt} = \frac{\gamma_e-1}{\gamma_i-1}\frac{T_i}{T_e}\left (  \frac{\delta_e}{\delta_i}-\frac{\tau_{\textrm{diss}}}{\tau_{\textrm{cool}}}\frac{u_e}{\delta_i(u_i+u_e)}\right )\frac{D\ln{T_i}}{Dt}\text{ .}
\end{equation}

Therefore, the difference in adiabatic indices between relativistic electrons and non-relativistic ions causes an effective reduction in the fluctuations of the electron temperature. This theoretically framework is in agreement with our results of $M_3$ distributions presented in Fig~\ref{fig:Light-curve_M3} and Table~\ref{tab:M3_freq}, and it aligns with our interpretation that the evolved $T_e$ in 2T simulations with variable adiabatic indices fluctuates less than the $T_e$ derived from the $R(\beta)$ prescription. 

\subsection{Comparison to semi-analytical models}
\label{sec:semi-analytical models}

In MAD states, we find substantial heating and synchrotron radiative cooling in areas with significant gas compression and high magnetisation. Our results for radiative efficiency align with those of \citet{Liska2024ApJ...966...47L}, and challenge the paradigm in ADAF models, with low $\eta_{\mathrm{rad}}$ across all radii. In our 2T and 2TC simulations, we obtain $ \left< T_e \right>\approx 10^{11} \text{K}$, time averaged between $[16-29]\times 10^3 r_g/c$ and measured at $10r_g$, consistent with the values reported in spectral fitting modelling  \citep[e.g.][]{Dexter2010ApJ...717.1092D,Markoff2001A&A...379L..13M,Yuan2002A&A...383..854Y} and EHT polarisation modelling  (\citealt{Dexter2016MNRAS.462..115D}; \citetalias{SgrAEHTC2017V,SgrAEHTC2017VIII}).

In the ADAF model by \cite{Narayan1995Natur.374..623N}, the radiative efficiency $\eta_{\mathrm{rad}}$ is less than 0.001 at all radii, meaning that less than $0.1\%$ of the available accretion energy is radiated, with the flow being advection-dominated. For $r < 200r_g$, ion-electron coupling weakens, causing $T_e$ to saturate, while $T_i$ continues to follow the virial temperature profile. They found that
$T_e$ rises up to $\sim 10^{10}\text{K}$ in accreting black hole flows. For \sgra, \citet{QuataertNarayan1999ApJ...520..298Q} achieved a good spectral fit using radiatively inefficient accretion flow (RIAF) models with $\delta_e = 0.55$ and required $T_e \sim 10^{11} \text{K}$ close to the BH, which is larger than in ADAF models. In semi-analytical jet models of \sgra, \citet{Markoff2001A&A...379L..13M} and \citet{Yuan2002A&A...383..854Y} obtained a good spectral fit with electron temperatures reaching $\sim 10^{11}\text{K}$.

Recently, the EHTC reported a large resolved polarisation fraction of 24–28\%, with a peak around 40\% \citepalias{SgrAEHTC2017VII}. Synchrotron emission is intrinsically highly polarised, but as light traverses a magnetised medium, it undergoes Faraday depolarisation \citep[e.g.][]{Ricarte2020MNRAS.498.5468R}. Faraday depth, which quantifies the rotation of polarised light, is inversely proportional to $T_e^2$ \citep{Jones1979ApJ...228..268J}. Consequently, higher $T_e$ leads to reduced depolarisation. In one-zone models, the total flux and optical depth constraints for \sgra\ indicate small Faraday depths, which account for the high polarisation fraction. Based on estimates of $n_e \approx 10^6 \, \mathrm{cm}^{-3}$ and $B \approx 29 \, \mathrm{G}$, it is expected that $T_e \approx 10^{11} \, \mathrm{K}$ at $\beta = 1$ (\citealt{Dexter2016MNRAS.462..115D}; \citetalias{SgrAEHTC2017V,SgrAEHTC2017VIII}), consistent with our results. Locally, $\beta_i$ can drop below unity, particularly at the disk-jet interface and within flux tubes. However, the time-averaged $ \left< \beta_i \right>\approx1$ in the disk at $\lesssim 20r_g$, as shown in Fig.~\ref{fig:2TC_vertical}.


\section{Conclusions}
\label{sec:Conclusions}

The EHTC has provided significant insights into the Galactic center source \sgra. Models based on 1T GRMHD simulations have been able to explain aspects of observations across radio to X-ray wavelengths in quiescent  \citepalias{SgrAEHTC2017V,SgrAEHTC2017VIII}, and flaring states \citep[e.g.][]{Chatterjee2021MNRAS.507.5281C,Scepi2022MNRAS.511.3536S}. However, none of the EHTC ray-traced models fully satisfy all the constraints drawn from multiwavelength observations at 86 GHz, 230 GHz, 2.2 $\mathrm{\mu m}$, and in the X-ray \citepalias{SgrAEHTC2017II,SgrAEHTC2017VII}. The 230 GHz variability constraint is particularly stringent, as nearly all EHTC models in a MAD state exhibit greater variability \citepalias{SgrAEHTC2017V} than historical observations (\citealt{Wielgus2022ApJ...930L..19W} and references therein). This limitation is possibly attributed to the modelled prescription of $T_i/T_e$, which is based on the local plasma magnetisation \citep[so called $R(\beta)$ prescriptions, e.g.][]{Moscibrodzka2016A&A...586A..38M,Anantua2020MNRAS.493.1404A}. In reality, $T_e$ is fundamentally influenced by microphysical plasma and radiation interactions, and does not depend trivially on $T_i$. A first-principles kinetic approach is required to model these collisionless effects \citep{Parfrey2019PhRvL.122c5101P,Crinquand2022PhRvL.129t5101C,Galishnikova2023PhRvL.130k5201G}.

Our investigation into 2T thermodynamics within MAD GRMHD simulations addresses some of these limitations by evolving both electron and ion temperatures \citep{Ressler2015MNRAS.454.1848R,Sadowski2017MNRAS.466..705S}. We do not resolve the actual heating mechanisms; instead, we assume that heating at the grid scale occurs through sub-grid magnetic reconnection \citep{Rowan2017ApJ...850...29R}. We do not model non-thermal electron distributions which are likely non-negligible \citep{Moscibrodzka2024arXiv241206492M}. We perform an analysis of variability that systematically adds one layer of physics at a time (see Table~\ref{tab:M3_freq}). By incorporating the 2T treatment, variable adiabatic index, and radiative cooling of electrons, we achieve a closer match to historical 228 GHz variability compared to 1T simulations. We find an effective reduction of nearly 50\% in the values of the three-hour modulation index ($M_3$) distribution. Additionally, we find that $M_3$ increases with frequency and does not significantly depend on the angle between the observer’s line of sight and the angular momentum vector of the accretion disk.


In a 2T GRMHD simulation, we extract two light curve predictions: one using the evolved $T_e$ directly from the simulation, and another where the evolved $T_e$ is ignored and instead calculated in post-processing with the $R(\beta)$ prescription. We find that $M_3$ is lower when using the evolved $T_e$, suggesting that the $T_e$ derived from the $R(\beta)$ prescription exhibits greater fluctuations (see Fig.~\ref{fig:Light-curve_M3}). This result is consistent with theoretical expectations for a 2T treatment, where the difference in adiabatic indices between relativistic electrons and non-relativistic ions causes an effective reduction in the fluctuations of the electron temperature (see section~\ref{sec:Fluctuations_Te_Ti}).

Even for the very low accretion rates relevant for \sgra\, ($f_{\mathrm{edd}} \approx 10^{-7}$), we find that radiative cooling of electrons—via synchrotron, inverse Compton, and bremsstrahlung processes—still affects the accretion flow, lowering $T_e$ in the inner $50r_g$ accretion disk (see Fig.~\ref{fig:disk_vs_r}), reducing average 228 GHz synchrotron flux, and shifting the $M_3$ distribution towards lower variability values by roughly 10\%. We find that synchrotron emission dominates over inverse Compton, while bremsstrahlung is negligible, consistent with \citet{Yoon2020MNRAS.499.3178Y}. Our estimated radiative efficiency, $\eta_{\mathrm{rad}} \approx [0.02-0.05]$, is broadly consistent with the value $\eta_{\mathrm{rad}} \approx 0.03$ obtained from GRRMHD MAD simulations that include radiative transport \citep{Liska2024ApJ...966...47L}. These results challenge the paradigm in ADAF models, with $\eta_{\mathrm{rad}}<0.001$ across all radii. MAD states have higher radiative efficiency, while in SANE states, cooling is expected to be less significant especially at low accretion rates. Therefore, electron radiative cooling is non negligible if the accretion flow of \sgra\, has a dynamically strong magnetic field, typical of a MAD state, as favoured by the current EHT observations.

Despite these improvements, further progress is required, as our 2T simulations still show more variability than historical observations of \sgra. For example, the more realistic stellar wind-fed accretion models better predict the submillimetre variability due to the comparatively lower levels of small-scale turbulence compared to SANE and MAD models \citep{Murchikova2022ApJ...932L..21M}. Additionally, our simulations do not resolve plasmoid-mediated magnetic reconnection, that requires higher resolution \citep{Salas2024MNRAS.533..254S} or an effective resistivity or viscosity. Magnetic reconnection can cause the dissipation of magnetic energy into heat \citep{Ripperda2019ApJS..244...10R, Ripperda2022ApJ...924L..32R}, thereby potentially influencing the radiative efficiency. Furthermore, ideal magnetohydrodynamics can never capture kinetic effects, like pressure anisotropy that can influence synchrotron emission and absorption \citep{Galishnikova2023PhRvL.130k5201G,Galishnikova2023ApJ...957..103G}. An inherent uncertainty in our approach stems from not resolving the heating mechanisms and relying on a single heating prescription. In future work we will explore heating prescriptions based on first-principles kinetic simulations in particular regions of the GRMHD domain, approximating a number of important subgrid effects of collisionless physics not captured by GRMHD simulations.


\section*{Acknowledgements}

We thank Charles Gammie and Andrew Chael for stimulating discussions. L.S and S.M. were supported by a Dutch Research Council (NWO) VICI award, grant No. 639.043.513 and by a European Research Council (ERC) Synergy Grant "BlackHolistic" grant No. 101071643. In addition, L.S. was supported by a Colfuturo Scholarship, in partnership with the Colombian Ministry of Science. ML was supported by the John Harvard, ITC and NASA Hubble Fellowship Program fellowships. K.C. was supported in part by grants from the Gordon and Betty Moore Foundation and the John Templeton Foundation to the Black Hole Initiative at Harvard University, and by NSF award OISE-1743747. G.M. was supported by a Canadian Institute of Theoretical Astrophysics (CITA) postdoctoral fellowship and by a Netherlands Research School for Astronomy (NOVA), Virtual Institute of Accretion (VIA) postdoctoral fellowship. GM acknowledges support from the Simons Collaboration on Extreme Electrodynamics of Compact Sources (SCEECS). O.P. acknowledges funding from VIA within NOVA. B.R. is supported by the Natural Sciences \& Engineering Research Council of Canada (NSERC) and by a grant from the Simons Foundation (MP-SCMPS-00001470). Research at the Flatiron Institute is supported by the Simons Foundation. This research was enabled by using resources from Calcul Quebec (http://www.calculquebec.ca) and Compute Canada (http://www.computecanada.ca). This work used the Dutch national e-infrastructure with the support of the SURF Cooperative using grant no. EINF-3036, EINF-5383 and EINF-9222, which is (partly) financed by the Dutch Research Council (NWO), for post-processing of simulation data.

\section*{Data Availability}

The simulation post-processed data used to plot the images in this work are available in Zenodo at \href{http://doi.org/10.5281/zenodo.14793884}{http://doi.org/10.5281/zenodo.14793884}.




\bibliographystyle{mnras}
\bibliography{main} 




\appendix

\section{Coulomb coupling}
\label{sec:Coulomb}

The transfer of energy between ions and electrons is characterised by the Coulomb coupling rate in the comoving frame, which in units of $\mathrm{erg\,cm^{-3}\,s^{-1}}$is equivalent to,
\begin{equation}
\begin{split}
    q^{C}=\frac{3m_e}{2m_i}\bar{n}n_e\ln &\Lambda\frac{ck\sigma_T\left ( T_i-T_e \right )}{K_2(1/\theta _i)K_2(1/\theta _e)} \\
    &\times \left[ \frac{2\left ( \theta_e + \theta_i \right )^2+1}{\theta_e + \theta_i}K_1(1/\theta _m)+2K_0(1/\theta _m) \right ]\text{ ,}
	\label{eq:Coulomb_coupling}
\end{split}
\end{equation}
with $\theta_m=\left ( 1/\theta _e+1/\theta _i \right )^{-1}$ and $m_i=m_p\left ( X+4Y \right )$. $K_i$ is the modified Bessel function of the $i$th order, $\ln \Lambda\approx20$ is the Coulomb logarithm, and $\bar{n}=\left ( X+Y \right )\rho/m_p$ is the number density \citep{Stepney1983MNRAS.204.1269S,Sadowski2017MNRAS.466..705S}. \cite{Ryan2017ApJ...844L..24R} suggested that Coulomb collisions become as important as viscous heating at $f_{\mathrm{edd}} \approx 10^{-4}$. \cite{Dexter2021ApJ...919L..20D} and \cite{Liska2024ApJ...966...47L} found Coulomb collisions are not important until $f_{\mathrm{edd}} \approx 10^{-3}$. Conversely, semi-analytical models typically assume that Coulomb collisions dominate for $f_{\mathrm{edd}} \gtrsim 10^{-2}$ \citep{Esin1997ApJ...489..865E}.



\bsp	
\label{lastpage}
\end{document}